\documentclass[journal]{IEEEtran}
\usepackage{epsfig,rotating,setspace,latexsym,amsmath,epsf,amssymb,bm,theorem}
\usepackage{cite,psfrag,bbm}
\usepackage{amsfonts,authblk,subfigure}
\usepackage{graphicx}

\newtheorem{Lemma}{Lemma}

\newenvironment{Proof}[1]{\medskip\par\noindent
{\bf Proof:\,}\,#1}{{\mbox{\,$\blacksquare$}\par}}

\def \n2{{N_0 \over 2}}

\def \h5{\hspace{0.5in}}

\begin{document}

	
	\title{Optimizing Information Freshness Through Computation-Transmission Tradeoff and Queue Management in Edge Computing}
	
	\author{Peng Zou \qquad Omur Ozel \qquad Suresh Subramaniam\thanks{The authors are with the Department of Electrical and Computer Engineering, George Washington University, Washington, DC 20052. Emails: \{pzou94, ozel, suresh\}@gwu.edu. Part of this work appears in the Proceedings of IEEE PIMRC, Istanbul, Turkey, September 2019 \cite{pimrc19}.}\vspace{-0.24in}}

	\maketitle 
	
	\begin{abstract}
		Edge computing applications typically require generated data to be preprocessed at the source and then transmitted to an edge server. In such cases, transmission time and preprocessing time are coupled, yielding a tradeoff between them to achieve the targeted objective. This paper presents analysis of such a system with the objective of optimizing freshness of received data at the edge server. We model this system as two queues in tandem whose service times are independent over time but the transmission service time is monotonically dependent on the computation service time in mean value. This dependence captures the natural decrease in transmission time due to lower offloaded computation. We analyze various queue management schemes in this tandem queue where the first queue has a single server, Poisson packet arrivals, general independent service and no extra buffer to save incoming status update packets. The second queue has a single server receiving packets from the first queue and service is memoryless. We consider the second queue in two forms: (i) No data buffer and (ii) One unit data buffer and last come first serve with discarding. We analyze various non-preemptive as well as preemptive cases. We perform stationary distribution analysis and obtain closed form expressions for average age of information (AoI) and average peak AoI. Our numerical results illustrate analytical findings on how computation and transmission times could be traded off to optimize AoI and reveal a consequent tradeoff between average AoI and average peak AoI. \vspace{0.05in} 
	\end{abstract}

		
	\section{Introduction}  

This paper is motivated by emerging edge computing applications in which generated data are preprocessed at the source and then transmitted to an edge server for further processing. In such a scenario, there is typically a tradeoff between the amount of preprocessing and the amount of data to be transmitted. Various Internet-of-Things (IoT) and edge computing applications require these tandem operations with examples spanning sensor networking, camera networks and vehicular communication networks. The data in these cases are representative forms of physical measurements such as sound, image, and temperature with dependence on location, system, and specific application. In such scenarios, it is critical to decide the portion of computation done on the source side with respect to those delegated to significantly more powerful\footnote{By the adjective powerful, we refer to computation capabilities, speed, and energy resources available to the server.} servers through communication\cite{aydin2018,opadere2019joint}. In this paper, we address this fundamental issue using computation and communication queues in tandem.

This paper focuses on freshness of information obtained at the end of computation and communication operations. A new metric termed age of information (AoI) has found considerable attention in the recent literature as a measure of freshness of available information. After the pioneering works \cite{kaul2012real,kaul2012status} that analyze queuing models motivated from vehicular status update systems, the AoI metric has been found useful in various scenarios. \cite{inoue2018general} provides a general AoI analysis in various preemptive and non-preemptive queuing disciplines coming after earlier works such as \cite{costa2016age, kam2018age, najm2016age}. References \cite{2018information, bacinoglu2015age, yates2015lazy, wu2017optimal_ieee, arafa2017age, bacinoglu2017scheduling, farazi2018average} consider AoI in energy harvesting communication systems. Evolution of AoI through multiple hops in networks has been characterized in \cite{bedewy2017age, talak2017minimizing, yates2018age, yates2018status, maatouk2018age}. \cite{Alabbasi2018JointIF} considers scheduling data flows in vehicular communication networks. More recently, \cite{Gong2019ReducingAF,xu2019peak} consider AoI analysis with tandem computing and communication queues. 

\begin{figure}[!t]
		\centering{
			\hspace{-0.4cm} 
			\includegraphics[totalheight=0.095\textheight]{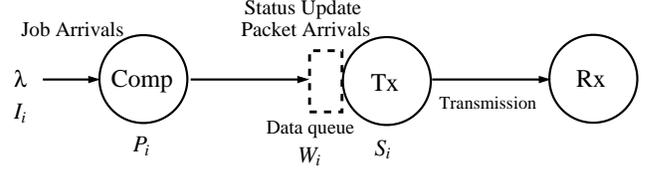}}\vspace{-0.1in}
		\caption{\sl System model with status update packets arriving to a single server transmission queue from the output of a computation server queue in tandem.}\vspace{-0.2in}
		\label{fig:1} 
	\end{figure}

	Our model is closely related to those considered in \cite{Gong2019ReducingAF,xu2019peak} modeling IoT edge computing scenarios. Our system involves tandem queues where a computation-type first queue determines status update packets to be sent to a powerful server having a remote monitoring receiver, as shown in Fig. \ref{fig:1} where the focus is on a single job server and a single transmission server. In our work, the computation time has a general distribution. In this model, we provide a general analysis with packet management for both average AoI and average peak AoI; and we optimize the tradeoff between computation and transmission times to maintain best AoI performance at the receiving end. The contributions of this paper are as follows:
	\begin{itemize}
	\item We provide a general analysis for the system in Fig. \ref{fig:1} with Poisson arrivals having rate $\lambda$ and processing time $P_i$ for a job having a general distribution. Moreover, a monotonic dependence is assumed between the mean service time of the first server and the second server which is typical of edge computing applications. Works in the literature on AoI through tandem queues (such as \cite{Gong2019ReducingAF,xu2019peak}) address this model in more specific forms.
	
	\item We  address four different schemes: two preemptive and two non-preemptive schemes. In the non-preemptive schemes, the first queue has no buffer and the second queue may or may not have a data buffer. In all cases, we obtain \textit{exact closed-form expressions} for both average peak AoI and average AoI providing explicit dependencies among the parameters in the system. To obtain these expressions, we use modified versions of equivalent queues that were first developed in our earlier works \cite{infocom_w, infocom_arxiv}. This is in contrast to works in the literature such as \cite{Gong2019ReducingAF,xu2019peak} that either provide approximate expressions or exact expressions under memoryless computation service for average AoI and average peak AoI separately. Additionally, these works consider only non-preemptive queues in tandem. Our work substantially extends these works in these aspects.
	
	\item Based on the monotonic dependence between mean computation and transmission times, we numerically determine the mean computation time that hits the best tradeoff to optimize a combination of average age of information and average peak age of information at the receiver. 
	     
	\item Our numerical results show the benefits obtained by judiciously determining the computation time in reducing AoI under various system settings including different computation time distributions and monotonic dependences between computation and transmission. In particular, our results make clear the contrast between AoI performances of non-preemptive and preemptive tandem queueing with respect to the variance of computation time (reminiscent of the one discussed in the seminal paper \cite{talak2018can} for single-server queues). Additionally, our results reveal a peculiar tradeoff between average peak AoI and average AoI generated by the tandem nature of the queueing system under dependent service times. 
	\end{itemize}
	\vspace{-0.1in}

	\section{The Tandem Queue Model}
	\label{sec:Model}
	
	We consider a system with a computation queue followed by a transmission queue as shown in Fig. \ref{fig:1}. In the sequel, we interchangeably refer to these queues as first and second queues. The computation jobs arrive according to a Poisson process with rate $\lambda$. The jobs enter the computation server only if it is idle and the aging process starts as soon as the job enters the server and the computation starts. Job computation times are identically distributed with a general distribution and independent over time and other parameters. As soon as the computation is completed, a status update packet, whose length is determined by the computation time, is sent to the transmission queue. The transmission queue includes a transmitter (Tx) and a receiver (Rx) where a monitor resides. There may be a single data buffer to save the latest arriving packet when transmission server is busy. Update packets are transmitted one at a time and transmission time is exponentially distributed with mean $\mu$.
	
	We analyze this tandem queue model by introducing packet management schemes in the transmission queue inspired by those in references \cite{costa2016age,inoue2018general}. We assume that there is no buffer in the first queue throughout the paper. On the other hand, the second queue may or may not have a buffer as reflected in Fig. \ref{fig:1}. Two packet management schemes in the second queue are addressed: GI/M/1/1 and GI/M/1/2* where the former involves zero data buffer whereas the latter involves one unit data buffer along with last come first serve with discarding (compatible with the usual notation of $2^*$ considered in the earlier literature \cite{costa2016age,inoue2018general}). We also address preemption and consider two schemes: M/GI/1 with preemption followed by GI/M/1$2^*$ and M/GI/1/1 followed by GI/M/1 with preemption queue. Note that in all cases the second queue has independent arrivals. The tandem nature of these queues renders the resulting problems new and the dependence in service times between them leads to novel problems that have not been analyzed before to the best of our knowledge. We will perform stationary distribution analysis for average age and average peak age. The time for a job to be served in the first queue has a general distribution $f_P(p)$, $p \geq 0$, independent of other system variables and independent over time. Corresponding to the general distribution, we have the moment generating function (MGF) evaluated at $-\gamma$ for $\gamma \geq 0$: \[M_{P}(\gamma) \triangleq \mathbb{E}[e^{-\gamma P}]. \] We also use $M_{(P,1)}(\gamma)$ and $M_{(P,2)}(\gamma)$ to denote, respectively, the first and second derivatives of $M_P$ at $- \gamma$.

	We let $t_i$ denote the time stamp of the event that job $i$ enters the computation server (we index only those that enter and exit the first server and assume no packet arrives when the computation server discards a packet), and $t_i'$ denote that of the event that the resulting packet $i$ (if selected for transmission) is delivered to the receiver. For each index $i$, there is a job and a packet. It is, however, to be noted that the age of the packet is determined with respect to the time its corresponding job enters the computation server.\footnote{This relativity is inherent in multi-hop systems and has been the topic of another work of ours in \cite{isit_arxiv}.} Instantaneous age of information (AoI) is the difference of current time and the latest time stamp at the receiver: \begin{align} \Delta(t)=t-u(t) \end{align} where $u(t)$ is the time stamp of the latest packet at the receiver at time $t$. We express $u(t)=t_{i^*}$ where $i^*=\max\{i: \ t_i' \leq t\}$.
	
	For all schemes, we use a corresponding equivalent queue model that yields an identical AoI pattern to our system's and that simplifies analysis. This approach first appeared in our earlier work \cite{infocom_w, infocom_arxiv} for a single-server queue and we adapt this approach for each scheme. Common to these equivalent models is that \emph{all} jobs/packets entering both the first queue and the second queue are served. Jobs that are discarded in the computation queue are not counted whereas those that enter the second queue but discarded there are counted. We next explain the equivalent queue models for each scheme. \vspace{-0.1in}

	\subsection{Equivalent Queue for Non-Preemptive Schemes}
	\label{sec:eqmodel1}
	
	In this section, we provide details of equivalent models for two non-preemptive schemes. 
	
\subsubsection{M/GI/1/1 followed by GI/M/1/1} In this equivalent model, all arriving packets to the second queue are stored in the queue, the data buffer capacity is unlimited and no packet is discarded. We allow multiple packets to be served at the same time in the second queue. An arriving packet may find the second queue in Idle (Id) or Busy (B) states. If a packet enters the second queue in state (Id), then that packet's service starts right away; otherwise, its service starts right after the current service period, and the idle period waited for the arrival of the next packet that enters service in the original system. The packets arriving to the second queue in state (B) are served together with all other packets that arrive during the same busy period. 
	
	\begin{figure}[!t]
		\centering{
			\hspace{-0.5cm} 
			\includegraphics[totalheight=0.21\textheight]{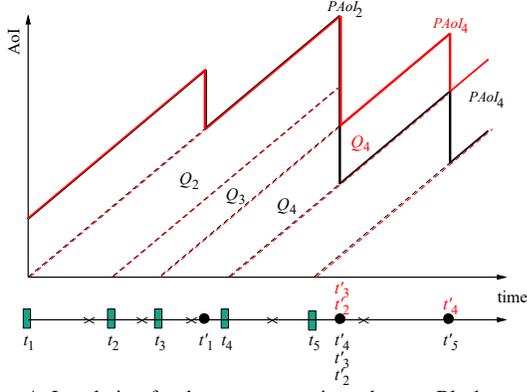}}\vspace{-0.15in}
		\caption{\sl AoI evolution for the non-preemptive schemes: Black one belongs to M/GI/1/1 followed by GI/M/1/1, and red one belongs to M/GI/1/1 followed by GI/M/1/$2^*$.}
		\label{fig:GM11} 
		\vspace{-0.2in}
	\end{figure}
	
	
	\subsubsection{M/GI/1/1 followed by GI/M/1/$2^*$}
		
	In this equivalent model, different from the earlier one, once a packet enters the second queue in state (B), that packet's service starts after the end of the current service period. The packets arriving to the second queue in state (B) are served together with all other packets that arrive during the same busy period. 
	
	Fig. \ref{fig:GM11} illustrates sample paths corresponding to the AoI under the equivalent tandem queue models for M/GI/1/1/ followed by GI/M/1/1 (shown in black color) as well as the one with GI/M/1/$2^*$ (shown in red color) in a comparative fashion using the same update arrival pattern. In the former scheme, at time $0$, job 1 enters the computation queue while both servers are idle and its computation time is marked as a cross in between $t_1$ and $t_2$. During the system time for packet $1$ in between $t_1$ and $t_1'$, jobs 2 and 3 arrive and their service in the computation queue end before the service of packet 1 in the transmission queue. Therefore, both packets enter the second queue and are kept in the buffer to be served later. The starting time for these packets' services constitutes the main difference between the two schemes represented by black and red schemes. In the red scheme, packets 2 and 3 are served immediately after $t_1'$ whereas in the black scheme, packets $2$ and $3$ are kept in the buffer until packet $4$ enters the second queue after which they are taken to service together. In the original schemes, packet $3$ replaces packet $2$ and it is served at $t_1'$ in the red scheme while packets $2$ and $3$ are discarded and the system is idle in the black scheme. At time $t_4$, job 4 enters the computation queue while both servers are idle in the black scheme whereas in the red scheme, the second server is busy when packet $4$ enters. In the red scheme, it waits in the buffer until packets $2$, $3$ are served at $t_2'$, $t_3'$ (coinciding in time axis). Packet $4$'s computation time is marked as the next cross after $t_4$. In the black scheme, once this computation is finished, packets $2$, $3$ and $4$ are served together and the points $t_2'$, $t_3'$, $t_4'$ coincide. In the actual system, only packet 4 is served. Note that in the illustration service time of packet 4 is assumed smaller in the black schemethan that in the red scheme.
	
	

	\subsection{Equivalent Queue for Preemptive Schemes}
	\label{sec:eqmodel2}
	
	In this section, we describe equivalent models for the two preemptive schemes. 

	\subsubsection{M/GI/1/1 followed by GI/M/1 with Preemption}
		
	In the equivalent queue for M/GI/1/1 followed by GI/M/1 with preemption, the main difference from its non-preemptive counterpart is that the system time for a packet entering the first queue has to take into account potential preemptions in the second queue. If, in the original system, it is discarded due to a new arrival during its service then that packet is assumed to be kept in the buffer until a successful transmission happens.

	\subsubsection{M/GI/1 with Preemption followed by GI/M/1/$2^*$}

	For the equivalent queuing model for M/GI/1 with preemption followed by GI/M/1/$2^*$, a major difference from other schemes is in the indexing of packets. We index incoming jobs that enter both the first and the second queues and those that cannot enter the first queue or the second queue are not counted. This causes a difference in the definition of inter-arrival times between preemptive and non-preemptive schemes. In particular, the inter-arrival times are the intervals between two arrivals that can enter both the first and the second queues. 
	
	\begin{figure}[!t]
		\centering{
			\hspace{-0.5cm} 
			\includegraphics[totalheight=0.21\textheight]{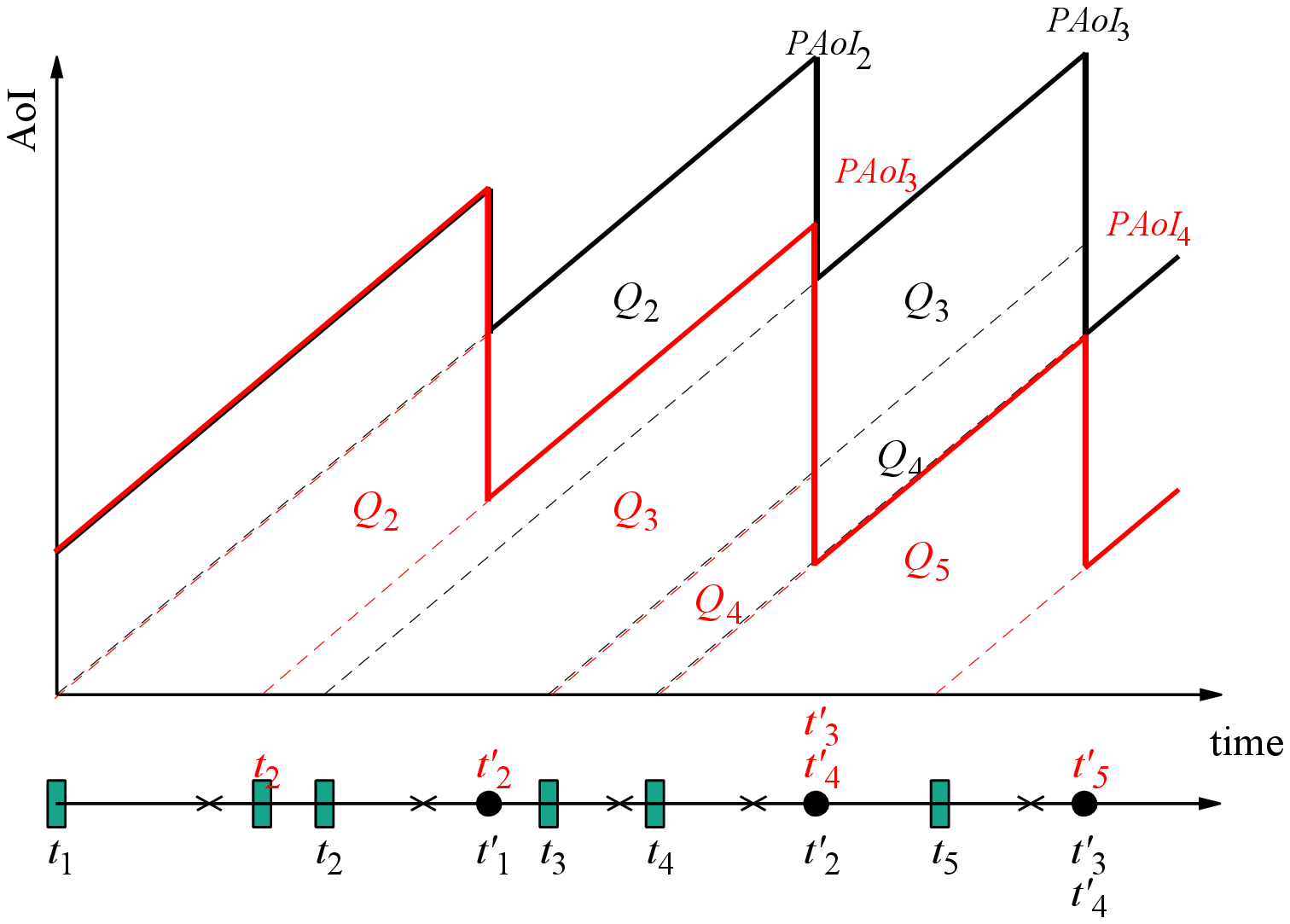}}\vspace{-0.15in}
		\caption{\sl AoI evolution for preemptive schemes: Red one belongs to M/GI/1/1 followed by GI/M/1 with preemption, and black one belongs to M/GI/1 with preemption followed by GI/M/1/$2^*$.}
		\label{fig:GM1P} 
		\vspace{-0.2in}
	\end{figure}

	We consider a sample path of the AoI for the equivalent queue models corresponding to the two preemptive schemes in Fig. \ref{fig:GM1P}. Here, the red scheme is M/GI/1/1 followed by GI/M/1 with preemption, and the black one is M/GI/1 with preemption followed by GI/M/1/$2^*$. At time $0$, job 1 enters the computation queue while the two servers are idle and its computation time ends at the cross in between $t_1$ and $t_2$ before $t_1'$. At $t_2$, job 2 arrives and finds both servers idle in the red scheme. In the black scheme, the packet that arrives at $t_2$ finds the first server busy and it is preempted by discarding the existing packet in service. In both black and red schemes, this computation ends at the time indicated as the next cross and the resulting update packet enters the second queue. In the red scheme, this packet is preempted by discarding the existing packet in service while in the black scheme, this packet is kept in buffer until the end of current service at $t_1'$. Then, job 3 arrives, its computation ends before job 4's arrival whose computation ends before the ongoing service in transmission server. In the red scheme, first packet 3 and then packet 4 are preempted by the transmission server. While packet 4 causes packet 3 to be discarded in the original system, in the equivalent model these two packets are served together. In the black scheme, both packet 3 and 4 are kept in the buffer of the second queue to be served after the end of the current transmission. Therefore, packets 3 and 4 are served together and the points $t_3'$, $t_4'$ coincide as in Fig. \ref{fig:GM1P} for both red and black schemes. Note that in the original system, only packet 4 is served. Packet 3 is discarded in both schemes.
	

	For all four tandem queuing models, we define the areas $Q_i$ under the triangular regions of the AoI curve as shown in Figs. \ref{fig:GM11}-\ref{fig:GM1P}. These definitions are identical to those in \cite{kaul2012real} for first come first serve queuing and we have the average AoI:
	\begin{equation}\label{aaoi}
	\mathbb{E}[\Delta]=\widetilde{\lambda}\left(\mathbb{E}[XT]+\frac{\mathbb{E}[X^{2}]}{2}\right),
	\end{equation}
	where $\widetilde{\lambda}$ is the effective arrival rate for the system defined as the rate of packets entering both the first and the second queues. Naturally, $\widetilde{\lambda}$ accounts for packets discarded in the first queue due to an arrival in a busy period. In M/GI/1 with preemption followed by GI/M/1/$2^*$, a packet in service in the first queue is unable to enter the second queue if a packet arrives during service time. Therefore only $\mbox{P}[S>I]=M_P(\lambda)$ (with $I$ representing inter-arrival time for first queue) portion of those that enter first queue can make it to the second queue and $\widetilde{\lambda}=\lambda M_P(\lambda)$. We also have the second moment of time $X$ between two successive arrivals that can enter the second queue for this scheme
	\begin{align}	
	\mathbb{E}[X^2]&=M_{(X,2)}(\gamma)|_{\gamma=0} =\frac{2-2\lambda M_{(P,1)}(\lambda)}{\lambda^2 (M_{P}(\gamma))^2}.
	\end{align}
	We evaluate the MGF of inter-arrival time $X$ in Section \ref{sec:eval3}. In other three schemes, we have the effective rate $\widetilde{\lambda} = \frac{\lambda}{ \lambda \mathbb{E}[P] + 1}$ as the packets arriving to the first queue while in service are discarded right away. Additionally, we have
	\begin{align}\label{ap1}
	\mathbb{E}[X^2]=\mathbb{E}[P^2] + \frac{2 \mathbb{E}[P]}{\lambda} + \frac{2}{\lambda^2}.
	\end{align}
	We also calculate average peak AoI. The peak AoI occurrences are shown in Figs. \ref{fig:GM11} and \ref{fig:GM1P}. In particular, $PAoI_{i^*}$ is the maximum $X_j+T_j$ among all packets $j$ served during a service period and $i^*$ is the smallest index among all of them. In Fig. \ref{fig:GM11}, packets $2$ and $3$ are served together and the peak AoI is $X_2 + T_2$. We assume the system is ergodic and we work with generic variables for inter-arrival time $X$, system time $T$ and $PAoI$ for the maximum $X_j + T_j$ among those that are served together. \vspace{-0.1in}
	
	\subsection{Functional Dependence of Mean Service Times}
	
	In our model, we assume that the mean service time of computation queue $\mathbb{E}[P]$ and mean service time of transmission queue $\mathbb{E}[S]=\frac{1}{\mu}$ are dependent through a monotone decreasing function $g$ as $\frac{1}{\mu} = g(\mathbb{E}[P])$. This dependence reflects the characteristic of computation server and the transmission server in terms of the time it takes to process jobs and packets, respectively. We are motivated by edge computing applications where some computation is performed at the transmitting device to reduce the amount of data to be transferred to a remote server. Under a fixed transmission rate, expected transmission time is proportional to the length of status update packet which is inversely proportional to the computation time. This operation could also be viewed as compression where the content of data transmission is reduced by removing the noise in the measurements partially or fully before transmission. Since a longer computation time leads to smaller packets, the function $g(.)$ is chosen to be monotone decreasing. \vspace{-0.1in}

	\section{M/GI/1/1 Followed by GI/M/1/1}
	\label{sec:eval1}
	
	In this scheme, the inter-arrival time between two successive jobs entering the computation queue (i.e., jobs $i-1$ and $i$) is $X_i$. Note that $X_i$ is independent over $i$ as $X_i = I_i + P_{i-1}$ where $I_i$ is exponentially distributed with mean $\frac{1}{\lambda}$. Therefore, we have the MGF of inter-arrivals as $M_{X}(\gamma) = \frac{\lambda}{\gamma + \lambda}M_{P}(\gamma)$.
	
	We let $T_i$ denote the system time for packet $i$ starting from its arrival to the computation queue until it is delivered to the receiver in the equivalent model. We have $T_i = P_i + W_i + I^*_i + S_i$ where $W_i \geq 0$ is the time packet $i$ spends in the second queue before entering service, $I^*_i$ is the additional idle period in second queue in case packet $i$ finds this queue busy and $S_i$ is the service time in the second queue. 
	
	Let us define the state of the second queue packet $i$ finds when it enters the queue as $K_i$, which can take (Id) and (B) states. We note that $K_i$ is a two-state Markov chain. Conditioned on $K_{i-1}=(Id)$, $K_{i}=(Id)$ only if $I_i + P_i > S_{i-1}$. Similarly, conditioned on $K_{i-1}=(B)$, $K_{i}=(Id)$ only if $I_i + P_i > W_{i-1}$ where $W_i$ denotes residual service time, which is also the waiting time for packet $i$ in the second queue conditioned on $K_i=(B)$. Note that both $W_i$ and $S_i$ are exponentially distributed with mean $\frac{1}{\mu}$ and they are independent variables. This generates a two-state Markov chain with transition probabilities:
	\begin{align*}
	\mbox{Pr}[K_{i}=(B)|K_{i-1}=(Id)]&=\mbox{Pr}[I_i + P_i < S_{i-1}]\\
	\mbox{Pr}[K_{i}=(Id)|K_{i-1}=(B)]&=\mbox{Pr}[I_i + P_i > W_{i-1}].
	\end{align*} 
Note that $\mbox{Pr}[K_{i}=(B)|K_{i-1}=(Id)] = 1- \mbox{Pr}[K_{i}=(Id)|K_{i-1}=(B)]$. So we just calculate
	\begin{align*}
	\mbox{Pr}[I_i + P_i < W_{i-1}]&=\mbox{Pr}[I_i + P_i < S_{i-1}] \\ &= \mathbb{E}[e^{-\mu (I_i + P_i)}] =\frac{\lambda}{\lambda + \mu} M_{P}(\mu). 
	\end{align*}
	Then, the stationary probabilities are $p_B=\mbox{Pr}[K_i = (B)]=\frac{\lambda}{\lambda + \mu} M_{P}(\mu)$ and $p_I = 1- p_B$.  \vspace{-0.1in}
	
	\subsection{Average AoI}
	
	In this subsection, we evaluate $\mathbb{E}[XT]$ and put it in (\ref{aaoi}) along with (\ref{ap1}) to get a closed form expression for average AoI. We next treat the two conditions $K_{i-1}=(Id)$ and $K_{i-1}=(B)$.
	
	\subsubsection{$\mathbb{E}[X_iT_i \ | \ K_{i-1}=(Id)]$}
	
	In this case, packet $i-1$ finds the second queue in (Id) state. $X_i = P_{i-1} + I_{i}$ and if $I_i + P_i > S_{i-1}$, then $T_i = P_i + S_i$. If $I_i + P_i < S_{i-1}$, then $T_i = P_i + W_i + I^*_i + S_i$ where $W_i$ is the residual service time observed by packet $i$ before entering the transmission server and $I^*_i$ is the idle period waited for the arrival of the next packet in second queue. Before going further, we state the following lemma for the expected value of $W_i + I^*_i$.
	
	\begin{Lemma}
		\label{Lem:W+I}
	In a GI/M/1/1 queue, let $G$ be a random variable representing inter-arrival time, $W$ be the residual service time for an arriving packet and $I^*$ be the idle period between the end of current service and next arrival. Then we have:
	\begin{equation}\label{lemma3}
	\mathbb{E}[W+I^*]=\frac{\mathbb{E}[G]}{\mbox{Pr}[K_{i}=(Id)|K_{i-1}=(B)]}
	\end{equation}
	\end{Lemma}
	
	\begin{Proof} Note that $W + I^*$ is a variable conditioned on $K_{i}=(B)$ which is the busy state of the second queue. It is equal to the time starting from the packet's arrival until the next packet that finds the second queue in (Id) state. Note that $K_{i}$ is a Markov chain and we have the transition probability from (B) to (Id). Now we can count the number of packets arriving while transitioning from (B) to (Id) and each time the queue remains in (B) state, we add an inter-arrival time to it. This is a geometric counting with mean value $\frac{1}{\mbox{Pr}[K_{i}=(Id)|K_{i-1}=(B)]}$ and the desired result follows from Wald's identity.
	\end{Proof}
	
	Now we evaluate the conditional expectation as:
	\begin{align*}
	&\mathbb{E}[X_iT_i|K_{i-1}=(Id)] = \mathbb{E}[(P_{i-1} + I_i)(P_i + S_i)] \\ &\hspace{1.3in} + \mathbb{E}[(P_{i-1} + I_i)(W_i+I^*_i) \mathbbm{1}_{I_i + P_i < S_i}] \\ &= \mathbb{E}[(P_{i-1} + I_i)(P_i + S_i)] \\ & \hspace{1.3in} + \mathbb{E}[(P_{i-1} + I_i)(W_i+I^*_i)e^{-\mu (I_i + P_i)}] \\ &= \mathbb{E}^2[P] + \frac{\mathbb{E}[P]}{\lambda} + \frac{1}{\lambda \mu} + \frac{\mathbb{E}[P]}{\mu} \\ & \hspace{0.9in} + \frac{(\lambda + \mu) \mathbb{E}[P] + 1}{(\lambda + \mu)}\frac{\lambda\mathbb{E}[P]+1}{(\lambda+\mu)-\lambda M_{P}(\mu)} M_{P}(\mu).
	\end{align*}

	\subsubsection{$\mathbb{E}[X_iT_i \ | \ K_{i-1}=(B)]$}
	
	In this case, packet $i-1$ finds the second queue in (B) state. $X_i = P_{i-1} + I_{i}$ and if $I_i + P_i > W_{i-1}$, then $T_i = P_i + S_i$. If $I_i + P_i < W_{i-1}$, then $T_i = P_i + W_i + I^*_i + S_i$. We evaluate the conditional expectation:
	\begin{align*}
	&\mathbb{E}[X_iT_i|K_{i-1}=(B)] = \mathbb{E}[(P_{i-1} + I_i)(P_i + S_i)] \\ &\hspace{1.2in} + \mathbb{E}[(P_{i-1} + I_i)(W_i+I^*_i) \mathbbm{1}_{I_i + P_i < W_{i-1}}] \\ &= \mathbb{E}[(P_{i-1} + I_i)(P_i + S_i)] \\ &\hspace{1.2in}+\mathbb{E}[(P_{i-1} + I_i)(W_i+I^*_i)e^{-\mu (I_i + P_i)}] \\ &= \mathbb{E}^2[P] + \frac{\mathbb{E}[P]}{\lambda} + \frac{1}{\lambda \mu} + \frac{\mathbb{E}[P]}{\mu} \\ &\hspace{0.8in}+ \frac{(\lambda + \mu) \mathbb{E}[P] + 1}{(\lambda + \mu)}\frac{\lambda\mathbb{E}[P]+1}{(\lambda+\mu)-\lambda M_{P}(\mu)} M_{P}(\mu).
	\end{align*}
	We finally use ergodicity of the system to get:
	\begin{align*}\nonumber
	\mathbb{E}[X_iT_i]&=\mathbb{E}[X_iT_i|K_{i-1}=(B)]p_{B}  +\mathbb{E}[X_iT_i|K_{i-1}=(Id)]p_{I} 
	 \\ &= \mathbb{E}^2[P] + \frac{\mathbb{E}[P]}{\lambda} + \frac{1}{\lambda \mu} + \frac{\mathbb{E}[P]}{\mu} \\ & \qquad + \frac{(\lambda + \mu) \mathbb{E}[P] + 1}{(\lambda + \mu)}\frac{\lambda\mathbb{E}[P]+1}{(\lambda+\mu)-\lambda M_{P}(\mu)} M_{P}(\mu),
	\end{align*} 
	where $p_B, p_I$ are the stationary busy and idle probabilities. \vspace{-0.1in}
	
	\subsection{Average Peak AoI}
	
	In this subsection, we evaluate $\mathbb{E}[X_{i^*} + T_{i^*}]$ where $i^*$ is the packet index corresponding to the minimum index in a given service period. We have $\mathbb{E}[X_{i^*}+T_{i^*}]=\frac{\mathbb{E}[(X_{i}+T_{i})\mathbbm{1}_{i=i^*}]}{\mbox{Pr}(i=i^*)}$ where $\mathbbm{1}_{i=i^*}$ is the indicator function of whether a given packet is the minimum index in a given service period and $\mbox{Pr}(i=i^*)$ refers to its probability. As before, we will treat the two conditions $K_{i-1}=(Id)$ and $K_{i-1}=(B)$ separately for both terms.

	\subsubsection{$\mathbb{E}[(X_{i}+T_{i})\mathbbm{1}_{i=i^*} \ | \ K_{i-1}=(Id)]$}
	
	In this case, if $I_i + P_i > S_{i-1}$, then $T_i = P_i + S_i$. If $I_i + P_i < S_{i-1}$, then $T_i = P_i + W_i + I^*_i + S_i$. Conditioned on (Id) state for packet $i-1$, the next packet index $i$ will be the minimum index among all those being served together with certainty. Therefore, we have $\mbox{Pr}(i=i^* \ | \ K_{i-1}=(Id)) = 1$. Additionally, we have
	\begin{align*}\nonumber
	\mathbb{E}[(X_{i}+&T_{i})\mathbbm{1}_{i=i^*} \ | \ K_{i-1}=(Id)]  \\ &\hspace{-0.5in}=  \mathbb{E}[P_{i-1} + I_i + P_i + S_i] +\mathbb{E}[(W_i + I^*_i) \mathbbm{1}_{I_i + P_i<S_i}] \\ &\hspace{-0.5in}=  \mathbb{E}[P_{i-1} + I_i + P_i + S_i] +\mathbb{E}[(W_i + I^*_i)e^{-\mu (I_i + P_i)}]  \\ &\hspace{-0.5in}=2\mathbb{E}[P] + \frac{1}{\lambda} + \frac{1}{\mu} + \frac{\lambda\mathbb{E}[P]+1}{\lambda + \mu-\lambda M_{P}(\mu)}M_{P}(\mu).
	\end{align*}

	\subsubsection{$\mathbb{E}[(X_{i}+T_{i})\mathbbm{1}_{i=i^*} \ | \ K_{i-1}=(B)]$}
	
	Conditioned on (B) state observed by packet $i-1$, the next packet index $i$ will be the minimum index among all those being served together only if the next packet $i$ arrives after the residual time $W_{i-1}$. Therefore, if $I_i + P_i < W_{i-1}$, then $\mathbbm{1}_{i=i^*} = 0$ and we have $\mbox{Pr}(i=i^* \ | \ K_{i-1}=(B)) = 1 - \frac{\lambda}{\lambda + \mu} M_{P}(\mu)$. In this case, if $I_i + P_i > W_{i-1}$, then $T_i = P_i + S_i$ and we have
	\begin{align*}
	\mathbb{E}[(X_{i}+T_{i})&\mathbbm{1}_{i=i^*} \ | \ K_{i-1}=(B)] \\ \nonumber &\hspace{-0.5in}= \mathbb{E}[(P_{i-1} + I_i + P_i + S_i)(1-e^{-\mu (I_i + P_i)})]   \\ &\hspace{-0.5in}=2\mathbb{E}[P] + \frac{1}{\lambda} + \frac{1}{\mu} -\frac{\lambda}{\lambda+\mu}M_{(P,1)}(\mu) \\&-(\mathbb{E}[P] + \frac{1}{\mu}+\frac{1}{\lambda+\mu})\frac{\lambda}{\lambda + \mu}M_{P}(\mu).  
	\end{align*}
	We use ergodicity and law of total expectation to conclude as:
	\begin{align*}\nonumber
	\mathbb{E}[(X_{i}+T_{i})&\mathbbm{1}_{i=i^*}] = 2\mathbb{E}[P] + \frac{1}{\lambda} + \frac{1}{\mu}  \\ &\hspace{-0.7in} + (1- p_B)\frac{\lambda\mathbb{E}[P]+1}{\lambda + \mu-\lambda M_{P}(\mu)}M_{P}(\mu) -\frac{\lambda p_B}{\lambda+\mu}M_{(P,1)}(\mu)\\&\hspace{-0.7in} - (\mathbb{E}[P] + \frac{1}{\mu}+\frac{1}{\lambda+\mu})\frac{\lambda p_B}{\lambda + \mu}M_{P}(\mu)\\
	\mbox{Pr}(i=i^*) &= 1 - p_B \frac{\lambda}{\lambda + \mu} M_{P}(\mu).
	\end{align*} \vspace{-0.05in}
	
	\section{M/GI/1/1 followed by GI/M/1/$2^*$}
	\label{sec:eval2}
	
	In the equivalent model for M/GI/1/1 followed by GI/M/1/$2^*$ scheme, we have $T_i = P_i + W_i + S_i$ where $W_i \geq 0$ denotes the length of time packet $i$ spends in the second queue before entering service and $S_i$ is the service time for packet $i$ in the second queue. Conditioned on $K_{i-1}=(Id)$, $K_{i}=(Id)$ only if $I_i + P_i > S_{i-1}$. Similarly, conditioned on $K_{i-1}=(B)$, $K_{i}=(Id)$ only if $I_i + P_i > W_{i-1} + S_{i-1}$ where $W_i$ denotes residual service time, which is also the waiting time for packet $i$ in the second queue conditioned on $K_i=(B)$. Note that both $W_i$ and $S_i$ are exponentially distributed with mean $\frac{1}{\mu}$ and they are independent variables. This generates a two-state Markov chain with transition probabilities:
	\begin{align*}
	\mbox{Pr}[K_{i}=(B)|K_{i-1}=(Id)]&=\mbox{Pr}[I_i + P_i < S_{i-1}]\\
	\mbox{Pr}[K_{i}=(Id)|K_{i-1}=(B)]&=\mbox{Pr}[I_i + P_i > W_{i-1} + S_{i-1}].
	\end{align*} 
	We calculate these probabilities as
	\begin{align*}
	\mbox{Pr}[I_i + P_i < S_{i-1}] &= \mathbb{E}[e^{-\mu (I_i + P_i)}] =\frac{\lambda}{\lambda + \mu} M_{P}(\mu), 
	\end{align*}
	\begin{align*}\nonumber
	\mbox{Pr}&[I_i + P_i < W_{i-1} + S_{i-1}] \\ &= \mathbb{E}[e^{-\mu (I_i + P_i)} + \mu (I_i + P_i)e^{-\mu (I_i + P_i)}] \\ &=  \frac{\lambda(\lambda + 2\mu)}{(\lambda + \mu)^2}M_{P}(\mu) + \frac{\lambda \mu}{\lambda + \mu} M_{(P,1)}(\mu), 
	\end{align*}
	and $\mbox{Pr}[I_i + P_i > W_{i-1} + S_{i-1}] = 1 - \mbox{Pr}[I_i + P_i < W_{i-1} + S_{i-1}]$. Then, the stationary probabilities are
	\begin{equation}\label{pb12}
	p_B=\frac{\lambda (\lambda + \mu) M_{P}(\mu)}{(\lambda+\mu)^2-\lambda \mu M_{P}(\mu) - \lambda \mu (\lambda + \mu) M_{(P,1)}(\mu)}, 
	\end{equation}
	and $p_I = 1- p_B$ where we define $p_B=\mbox{Pr}[K_i = (B)]$.  
	
	\subsection{Average AoI}
	
	In this subsection, we evaluate $\mathbb{E}[XT]$ and put it in (\ref{aaoi}) along with (\ref{ap1}) to get a closed form expression for average AoI. We next treat the two conditions $K_{i-1}=(Id)$ and $K_{i-1}=(B)$.
	
	\subsubsection{$\mathbb{E}[X_iT_i \ | \ K_{i-1}=(Id)]$}
	
	In this case, packet $i-1$ finds the second queue in (Id) state. $X_i = P_{i-1} + I_{i}$ and if $I_i + P_i > S_{i-1}$, then $T_i = P_i + S_i$. If $I_i + P_i < S_{i-1}$, then $T_i = P_i + W_i + S_i$ where $W_i$ is the residual service time observed by packet $i$ before entering the transmission server. We evaluate the conditional expectation as:
	\begin{align*}
	&\mathbb{E}[X_iT_i|K_{i-1}=(Id)] \\ &\quad = \mathbb{E}[(P_{i-1} + I_i)(P_i + S_i)]   +\mathbb{E}[(P_{i-1} + I_i)W_i \mathbbm{1}_{I_i + P_i < S_i}] \\ &\quad = \mathbb{E}[(P_{i-1} + I_i)(P_i + S_i)] +\mathbb{E}[(P_{i-1} + I_i)W_ie^{-\mu (I_i + P_i)}] \\ &\quad = \mathbb{E}^2[P] + \frac{\mathbb{E}[P]}{\lambda} + \frac{1}{\lambda \mu} + \frac{\mathbb{E}[P]}{\mu} + \frac{\lambda (\lambda + \mu) \mathbb{E}[P] + \lambda}{\mu (\lambda + \mu)^2} M_{P}(\mu).
	\end{align*}

	\subsubsection{$\mathbb{E}[X_iT_i \ | \ K_{i-1}=(B)]$}
	
	In this case, packet $i-1$ finds the second queue in (B) state. $X_i = P_{i-1} + I_{i}$ and if $I_i + P_i > W_{i-1} + S_{i-1}$, then $T_i = P_i + S_i$. If $I_i + P_i < W_{i-1} + S_{i-1}$, then $T_i = P_i + W_i + S_i$. We evaluate the conditional expectation:
	\begin{align*}
	&\mathbb{E}[X_iT_i|K_{i-1}=(B)] \\ & = \mathbb{E}[(P_{i-1} + I_i)(P_i + S_i)] +\mathbb{E}[(P_{i-1} + I_i)W_ie^{-\mu (I_i + P_i)}]\\ & + \mathbb{E}[(P_{i-1} + I_i)W_i\mu (I_i + P_i) e^{-\mu (I_i + P_i)}] \\ \nonumber &= \mathbb{E}^2[P] + \frac{\mathbb{E}[P]}{\lambda} + \frac{1}{\lambda \mu} + \frac{\mathbb{E}[P]}{\mu} + \frac{\lambda (\lambda + \mu) \mathbb{E}[P] + \lambda}{\mu (\lambda + \mu)^2} M_{P}(\mu)\\&+\frac{\lambda \mathbb{E}[P](\lambda + \mu + 1)}{(\lambda + \mu)^2}M_{P}(\mu) +\frac{2 \lambda M_{P}(\mu) + \lambda(\lambda + \mu)M_{(P,1)}(\mu)}{(\lambda + \mu)^3}.
	\end{align*}
	We finally use ergodicity of the system to get:
	\begin{align*}
	&\mathbb{E}[X_iT_i]\\&=\mathbb{E}[X_iT_i|K_{i-1}=(B)]p_{B} + \mathbb{E}[X_iT_i|K_{i-1}=(Id)]p_{I} \\
	 &= \mathbb{E}^2[P] + \frac{\mathbb{E}[P]}{\lambda} + \frac{1}{\lambda \mu} + \frac{\mathbb{E}[P]}{\mu} + \frac{\lambda (\lambda + \mu) \mathbb{E}[P] + \lambda}{\mu (\lambda + \mu)^2} M_{P}(\mu) \\ &\quad +p_B \frac{\lambda \mathbb{E}[P](\lambda + \mu) + 2\lambda}{(\lambda + \mu)^3}M_{P}(\mu) +p_B\frac{ \lambda \mathbb{E}[P] + 1}{\lambda + \mu}M_{(P,1)}(\mu),
	\end{align*} 
	where $p_B$ is as in (\ref{pb12}).
	
	\subsection{Average Peak AoI}
	
	In this subsection, we evaluate $\mathbb{E}[X_{i^*} + T_{i^*}]$ where $i^*$ is the packet index corresponding to the minimum index in a given service period. We have $\mathbb{E}[X_{i^*}+T_{i^*}]=\frac{\mathbb{E}[(X_{i}+T_{i})\mathbbm{1}_{i=i^*}]}{\mbox{Pr}(i=i^*)}$ where $\mathbbm{1}_{i=i^*}$ is the indicator function of whether a given packet is the minimum index in a given service period and $\mbox{Pr}(i=i^*)$ refers to its probability. As before, we will treat two conditions $K_{i-1}=(Id)$ and $K_{i-1}=(B)$ separately for both terms.

	\subsubsection{$\mathbb{E}[(X_{i}+T_{i})\mathbbm{1}_{i=i^*} \ | \ K_{i-1}=(Id)]$}
	
	In this case, if $I_i + P_i > S_{i-1}$, then $T_i = P_i + S_i$. If $I_i + P_i < S_{i-1}$, then $T_i = P_i + W_i + S_i$. Conditioned on (Id) state for packet $i-1$, the next packet index $i$ will be the minimum index among all those being served together with certainty. Therefore, we have $\mbox{Pr}(i=i^* \ | \ K_{i-1}=(Id)) = 1$. Additionally, we have
	\begin{align*}\nonumber
	\mathbb{E}[(X_{i}+&T_{i})\mathbbm{1}_{i=i^*} \ | \ K_{i-1}=(Id)]  \\ &\hspace{-0.5in}=  \mathbb{E}[P_{i-1} + I_i + P_i + S_i] +\mathbb{E}[W_i \mathbbm{1}_{I_i + P_i<S_i}] \\ &\hspace{-0.5in}=  \mathbb{E}[P_{i-1} + I_i + P_i + S_i] +\mathbb{E}[W_ie^{-\mu (I_i + P_i)}]  \\ &\hspace{-0.5in}=2\mathbb{E}[P] + \frac{1}{\lambda} + \frac{1}{\mu} + \frac{\lambda}{\mu(\lambda + \mu)}M_{P}(\mu).
	\end{align*}

	\subsubsection{$\mathbb{E}[(X_{i}+T_{i})\mathbbm{1}_{i=i^*} \ | \ K_{i-1}=(B)]$}
	
	Conditioned on (B) state observed by packet $i-1$, the next packet index $i$ will be the minimum index among all those being served together only if the next packet $i$ arrives after the residual time $W_{i-1}$. Therefore, if $I_i + P_i < W_{i-1}$, then $\mathbbm{1}_{i=i^*} = 0$ and we have $\mbox{Pr}(i=i^* \ | \ K_{i-1}=(B)) = 1 - \frac{\lambda}{\lambda + \mu} M_{P}(\mu)$. In this case, if $I_i + P_i > W_{i-1} + S_{i-1}$, then $T_i = P_i + S_i$. If $W_{i-1} < I_i + P_i < W_{i-1} + S_{i-1}$, then $T_i = P_i + W_i + S_i$ and we have
	\begin{align*}
	\mathbb{E}[(X_{i}&+T_{i})\mathbbm{1}_{i=i^*} \ | \ K_{i-1}=(B)] \\ & = \mathbb{E}[(P_{i-1} + I_i + P_i + S_i)(1-e^{-\mu (I_i + P_i)})] \\ &\ \ +\mathbb{E}[W_i \mu (I_i + P_i)e^{-\mu (I_i + P_i)}]  \\ &=2\mathbb{E}[P] + \frac{1}{\lambda} + \frac{1}{\mu} - \frac{(\mathbb{E}[P] + \frac{1}{\mu})\lambda}{\lambda + \mu}M_{P}(\mu). 
	\end{align*}
	We use ergodicity of the system to conclude as follows:
	\begin{align*}
	\mathbb{E}[(X_{i}+T_{i})\mathbbm{1}_{i=i^*}] &= 2\mathbb{E}[P] + (1- 2p_B)\frac{\lambda M_{P}(\mu)}{\mu(\lambda + \mu)} \\ & \qquad - p_B \frac{\mathbb{E}[P] \lambda}{\lambda + \mu}M_{P}(\mu) + \frac{1}{\lambda} + \frac{1}{\mu} \\
	\mbox{Pr}(i=i^*) &= 1 - p_B \frac{\lambda}{\lambda + \mu} M_{P}(\mu),
	\end{align*}
	and finally we have $\mathbb{E}[X_{i^*} + T_{i^*}] = \frac{\mathbb{E}[(X_{i}+T_{i})\mathbbm{1}_{i=i^*}] }{\mbox{Pr}(i=i^*)}$.
	
	\section{M/GI/1/1 followed by GI/M/1 with Preemption}
	\label{sec:eval4}
	
	In the equivalent preemption scheme, we have inter-arrival time $X_i = I_i + P_{i-1}$ as in earlier non-preemptive schemes. Different from earlier schemes, preemption determines the system time. Every task spends a waiting period $W'_i$ in buffer where $W'_{i}=\sum_{j=1}^{M}X'_{j}$ and $X'_{j}$ is $X_i$ provided that the current service time lasts longer and $M$ is a geometric random variable. So we have $T_i = P_i + W'_i + S'_i$ where $S'_i$ is the transmission time provided that current service end earlier than next arrival. 
	
	Now let us consider the M/GI/1/1 followed by GI/M/1 with preemption system. Conditioned on $K_{i-1}=(Id)$, $K_{i}=(Id)$ only if $I_i + P_i > S_{i-1}$. Similarly, conditioned on $K_{i-1}=(B)$, $K_{i}=(Id)$ only if $I_i + P_i > S_{i-1}$. $S_i$ are exponentially distributed with mean $\frac{1}{\mu}$. This generates a two-state Markov chain with transition probabilities:
	\begin{align*}
	\mbox{Pr}[K_{i}=(B)|K_{i-1}=(Id)]&=\mbox{Pr}[I_i + P_i < S_{i-1}]\\
	\mbox{Pr}[K_{i}=(Id)|K_{i-1}=(B)]&=\mbox{Pr}[I_i + P_i > S_{i-1}].
	\end{align*} 
	We calculate these probabilities as
	\begin{align*}
	\mbox{Pr}[I_i + P_i < S_{i-1}] &= \mathbb{E}[e^{-\mu (I_i + P_i)}] =\frac{\lambda}{\lambda + \mu} M_{P}(\mu),
	\end{align*}
	and $\mbox{Pr}[I_i + P_i > W_{i-1}] = 1 - \mbox{Pr}[I_i + P_i < W_{i-1}]$. Then, the stationary probabilities are $p_B=\mbox{Pr}[K_i = (B)]=\frac{\lambda}{\lambda + \mu} M_{P}(\mu)$ and $p_I = 1- p_B$.   
	
	\subsection{Average AoI}
	
	Now, we evaluate $\mathbb{E}[XT]$ and put it in (\ref{aaoi}) along with (\ref{ap1}) to get an expression for average AoI. We have $X_i=I_i+P_{i-1}$, $T_{i}=P_{i}+W'_{i}+S'_{i}$ and $\mathbb{E}[X_iT_i]$ is:
	\begin{align*}
	\mathbb{E}[X_iT_i]&=\mathbb{E}[(I_i+P_{i-1})(P_{i}+W'_{i}+S'_{i})]\\ &=(\frac{1}{\lambda}+\mathbb{E}[P])(\mathbb{E}[P]+\mathbb{E}[W']+\mathbb{E}[S']).
	\end{align*}
	 Closed form expressions for $\mathbb{E}[W']$ and $\mathbb{E}[S']$ are in Appendix \ref{App:extGM1P}.

	\subsection{Average Peak AoI}
	
	In this subsection, we evaluate $\mathbb{E}[X_{i^*} + T_{i^*}]$ where $i^*$ is the packet index corresponding to the minimum index in a given service period.  We have $\mathbb{E}[X_{i^*}+T_{i^*}]=\frac{\mathbb{E}[(X_{i}+T_{i})\mathbbm{1}_{i=i^*}]}{\mbox{Pr}(i=i^*)}$ where $\mathbbm{1}_{i=i^*}$ is the indicator function of whether a given packet is the minimum index in a given service period and $\mbox{Pr}(i=i^*)$ refers to its probability. As before, we will treat the two conditions $K_{i-1}=(Id)$ and $K_{i-1}=(B)$ separately for both terms.
	\subsubsection{$\mathbb{E}[(X_{i}+T_{i})\mathbbm{1}_{i=i^*} \ | \ K_{i-1}=(Id)]$}

	Conditioned on (Id) state observed by packet $i-1$, the next packet index $i$ will be the minimum index among all those being served together only if the next packet $i$ arrives after the service time $S_{i-1}$. Therefore, if $I_i + P_i < S_{i-1}$, then $\mathbbm{1}_{i=i^*} = 0$ and we have $\mbox{Pr}(i=i^* \ | \ K_{i-1}=(Id)) = 1 - \frac{\lambda}{\lambda + \mu} M_{P}(\mu)$. In this case, if $I_i + P_i > S_{i-1}$, then $T_i = P_{i}+W'_{i}+S'_{i}$ and we have
	\begin{align*}\nonumber
	&\mathbb{E}[(X_{i}+T_{i})\mathbbm{1}_{i=i^*} \ | \ K_{i-1}=(Id)]  \\ &=  \mathbb{E}[(P_{i-1} + I_i + P_i + W'_{i}+S'_{i})\mathbbm{1}_{I_i + P_i>S_{i-1}}] \\ &=  \mathbb{E}[(P_{i-1} + I_i + P_i + W'_{i}+S'_{i})(1-e^{-\mu (I_i + P_i)})]  \\ &=2\mathbb{E}[P]+\mathbb{E}[S'] -(\mathbb{E}[P] + \mathbb{E}[W']+\mathbb{E}[S'])\frac{\lambda}{\lambda + \mu} M_{P}(\mu)\\ &\quad -\frac{\lambda}{\lambda + \mu} M_{(P,1)}(\mu)-\frac{\lambda}{(\lambda + \mu)^2} M_{P}(\mu) + \frac{1}{\lambda} + \mathbb{E}[W'].
	\end{align*}

\subsubsection{$\mathbb{E}[(X_{i}+T_{i})\mathbbm{1}_{i=i^*} \ | \ K_{i-1}=(B)]$}

	Conditioned on (B) state observed by packet $i-1$, the next packet index $i$ will be the minimum index among all those being served together only if the next packet $i$ arrives after the service time $S_{i-1}$. Therefore, if $I_i + P_i < S_{i-1}$, then $\mathbbm{1}_{i=i^*} = 0$ and we have $\mbox{Pr}(i=i^* \ | \ K_{i-1}=(B)) = 1 - \frac{\lambda}{\lambda + \mu} M_{P}(\mu)$. In this case, if $I_i + P_i > S_{i-1}$, then $T_i = P_{i}+W'_{i}+S'_{i}$ and we have
	\begin{align*}\nonumber
	&\mathbb{E}[(X_{i}+T_{i})\mathbbm{1}_{i=i^*} \ | \ K_{i-1}=(B)] \\ &=  \mathbb{E}[(P_{i-1} + I_i + P_i + W'_{i}+S'_{i})\mathbbm{1}_{I_i + P_i>S_{i-1}}] \\ &=  \mathbb{E}[(P_{i-1} + I_i + P_i + W'_{i}+S'_{i})(1-e^{-\mu (I_i + P_i)})]  \\ &=2\mathbb{E}[P] + \frac{1}{\lambda} + \mathbb{E}[W']+\mathbb{E}[S'] -\frac{\lambda}{\lambda + \mu} M_{(P,1)}(\mu) \\ &\ -(\mathbb{E}[P] + \mathbb{E}[W']+\mathbb{E}[S'])\frac{\lambda}{\lambda + \mu} M_{P}(\mu) -\frac{\lambda}{(\lambda + \mu)^2} M_{P}(\mu).
	\end{align*}
	
	\noindent We use law of total expectation to conclude:
	\begin{align*}\nonumber
	&\hspace{-0.3in}\mathbb{E}[(X_{i}+T_{i})\mathbbm{1}_{i=i^*}] =2\mathbb{E}[P] + \frac{1}{\lambda} + \mathbb{E}[W']+\mathbb{E}[S']\\ &-(\mathbb{E}[P] + \mathbb{E}[W']+\mathbb{E}[S'])\frac{\lambda}{\lambda + \mu} M_{P}(\mu) \\ &-\frac{\lambda}{\lambda + \mu} M_{(P,1)}(\mu)-\frac{\lambda}{(\lambda + \mu)^2} M_{P}(\mu)
	 \\
	&\mbox{Pr}(i=i^*) = 1 - \frac{\lambda}{\lambda + \mu} M_{P}(\mu).
	\end{align*}

	\section{M/GI/1 with Preemption followed by GI/M/1/$2^*$}
	\label{sec:eval3}
	
	In this scheme, we analyze the inter-arrival time $X_i$ between two successive jobs that enter both the first and the second queues. We index them as jobs $i-1$ and $i$. Note that $X_i$ is the sum of three random variables: (a) idle time for the next arrival $I_i$, (b) a series of inter-arrival times $X'_{j}$ for all jobs discarded due to new arrivals before their processing times, and (c) processing time of job $i-1$ in the computation queue $P'_{i-1}$. That is, $X_i = I_i + P'_{i-1} +Y_i$. Here, $I_i$ is independent memoryless exponentially distributed with rate $\lambda$; $Y_i=\sum_{j=1}^{M}X'_{j}$ is the time spent in the processing period before the first successful processing happens and $M$ is a geometric random variable. Finally, $P'_{i-1}$ is distributed as $f_P(p)$ conditioned on $P$ being longer than the time until arrival of the next job. Therefore, we have $M_{X}(\gamma)= \frac{\lambda}{\gamma + \lambda}M_{P'}(\gamma)M_{Y}(\gamma)$.
	
Inspired from \cite[Lemmas 2, 3]{najm2016age}, we claim the following results and then compute the MGF of $Y_i$:
	\begin{Lemma}
		\label{Lem:G<F}
	Let G be a random variable and F be another one exponentially distributed with rate $\lambda$ independent of G. Conditioned on $\{G<F\}$, MGF of G is:
	\begin{equation}\label{lemma1}
	M_{(G|G<F)}(\gamma)=\frac{M_{G}(\gamma+\lambda)}{M_{G}(\lambda)}.
	\end{equation}	
	\end{Lemma}
\begin{Proof} We use the relation $f_{G|G<F}=f_{G}(t)\frac{\mathbb{P}(F>t)}{\mathbb{P}(G<F)}$ from \cite{najm2016age}. Since $\mathbb{P}(G<F)=M_{G}(\lambda)$ and $\mathbb{P}(F>t)=e^{-t\lambda}$, we have:
	\begin{align*}
	M_{(G|G<F)}(\gamma)&=\int_{0}^{+\infty}e^{-\gamma t}f_{G|G<F}\ \mathrm {d}t\\
	&=\int_{0}^{+\infty}e^{-\gamma t}f_{G}(t)\frac{e^{-t\lambda}}{M_{G}(\lambda)}\ \mathrm {d}t\\
	&=\frac{M_{G}(\gamma+\lambda)}{M_{G}(\lambda)}.
	\end{align*}
	\end{Proof}
	\begin{Lemma}
		\label{Lem:F<G}
		Let G be a random variable and F be another one exponentially distributed with rate $\lambda$ independent of G. Conditioned on $\{F<G\}$, MGF of F is:
		\begin{equation}\label{lemma2}
		M_{(F|F<G)}(\gamma)=\frac{1}{1-M_{G}(\lambda)}(\frac{\lambda}{\lambda+\gamma}-\frac{\lambda M_G(\gamma+\lambda)}{\lambda+\gamma}).
		\end{equation}
		
	\end{Lemma}
	\begin{Proof}
	We use the relation $f_{F|F<G}=f_{F}(t)\frac{\mathbb{P}(G>t)}{\mathbb{P}(F<G)}$ from \cite{najm2016age}. Since $\mathbb{P}(F<G)=1-M_{G}(\lambda)$ and $\mathbb{P}(G>t)=1-\mathbb{P}(G<t)$:
		\begin{align*}
		M_{(F|F<G)}(\gamma)&=\int_{0}^{+\infty}e^{-\gamma t}f_{F|F<G}\ \mathrm {d}t\\
		&=\int_{0}^{+\infty}e^{-\gamma t}f_{F}(t)\frac{1-\mathbb{P}(G<t)}{1-M_{G}(\lambda)}\ \mathrm {d}t\\
		&=\frac{1}{1-M_{G}(\lambda)}\left(\frac{\lambda}{\lambda+\gamma}-\frac{\lambda M_G(\gamma+\lambda)}{\lambda+\gamma}\right).
		\end{align*}
	\end{Proof}
	
	\noindent Now we compute the moment generating function of $X_i$. The moment generating function of $P'_{i-1}$ from Lemma \ref{Lem:G<F} is:
	\begin{equation*}
	M_{P'}(\gamma)=\frac{M_{P}(\gamma+\lambda)}{M_{P}(\lambda)}.
	\end{equation*}
	
	\noindent Then the moment generating function of $Y_{i}$ from (\ref{lemma2}) and \cite{najm2016age}:
	\begin{align*}
	M_{X'}(\gamma)&=\frac{1}{1-M_{P}(\lambda)}(\frac{\lambda}{\lambda+\gamma}-\frac{\lambda M_{P}(\gamma+\lambda)}{\lambda+\gamma}),\\
	M_Y(\gamma)&=\frac{M_P(\lambda)}{1-(1-M_P(\lambda))M_{X'}(\gamma)}\\
	&=\frac{(\lambda+\gamma)M_{P}(\lambda)}{\gamma+\lambda M_{P}(\gamma+\lambda)}.
	\end{align*}
	We derive expressions for $M_{(P',1)}(\gamma)$, $M_{(Y,1)}(\gamma)$ and $M_{(Y,2)}(\gamma)$ in Appendix \ref{App:extMGFp1MGFy1}. The moment generating function of the inter-arrival time $X_i$ is:
	\begin{align}\label{mgfx}
 M_{X}(\gamma)&= \frac{\lambda}{\gamma + \lambda}M_{P'}(\gamma)M_{Y}(\gamma)=\frac{\lambda M_{P}(\gamma+\lambda)}{\gamma+\lambda M_P(\gamma+\lambda)},
	\end{align} \begin{align} \nonumber
		M_{(X,1)}(\gamma)&=\frac{\lambda(M_{P}(\gamma+\lambda)+\gamma M_{(P,1)}(\gamma+\lambda))}{(\gamma+\lambda M_P(\gamma+\lambda))^2},\\
		M_{(X,2)}(\gamma)&=\frac{\lambda \gamma M_{(P,2)}(\gamma+\lambda)(\gamma+\lambda M_{P}(\gamma+\lambda))}{(\gamma+\lambda M_P(\gamma+\lambda))^3}\nonumber \\ &\qquad +\frac{\gamma M_{(P,1)}(\gamma+\lambda))(-1+\lambda M_{(P,1)}(\gamma+\lambda)}{(\gamma+\lambda M_P(\gamma+\lambda))^3} \nonumber \\ &\qquad -2\frac{\lambda(M_{P}(\gamma+\lambda))}{(\gamma+\lambda M_P(\gamma+\lambda))^3}. \label{mgfx12}
	\end{align}
	
	The second queue remains the same equivalent queue for GI/M/1/$2^*$ queue and in GI/M/1/$2^*$ scheme, we have $T_i = P_i + W_i + S_i$ where $W_i \geq 0$ denotes the length of time packet $i$ spends in the second queue before entering service and $S_i$ is the service time for packet $i$ in the second queue. 
	
	Now let us consider the M/GI/1 with preemption followed by GI/M/1/$2^*$ system. Conditioned on $K_{i-1}=(Id)$, $K_{i}=(Id)$ only if $I_i + P'_{i} +Y_i > S_{i-1}$. Similarly, conditioned on $K_{i-1}=(B)$, $K_{i}=(Id)$ only if $I_i + P'_{i} +Y_i > W_{i-1} + S_{i-1}$ where $W_i$ denotes residual service time, which is also the waiting time for packet $i$ in the second queue conditioned on $K_i=(B)$. Note that both $W_i$ and $S_i$ are exponentially distributed with mean $\frac{1}{\mu}$ and they are independent variables. This generates a two-state Markov chain with transition probabilities:
	\begin{align*}
	\mbox{Pr}[K_{i}=(B)|K_{i-1}=(Id)]&=\mbox{Pr}[I_i + P'_{i} +Y_i < S_{i-1}],\\
	\mbox{Pr}[K_{i}=(Id)|K_{i-1}=(B)] \\ =\mbox{Pr}[I_i &+ P'_{i} +Y_i> W_{i-1} + S_{i-1}].
	\end{align*} 
	We calculate these probabilities as
	\begin{align*}
	\mbox{Pr}[I_i + P'_{i} +Y_i< S_{i-1}] &= \mathbb{E}[e^{-\mu (I_i + P'_{i} +Y_i)}],
	\end{align*}
	\begin{align*}\nonumber
	&\mbox{Pr}[I_i + P'_{i} +Y_i < W_{i-1} + S_{i-1}] \\ &\quad =\mathbb{E}[e^{-\mu (I_i + P'_{i} +Y_i)} + \mu(I_i + P'_{i} +Y_i)e^{-\mu(I_i + P'_{i} +Y_i)}]. 
	\end{align*}
	Since $P'_{i}$ is independent of $P'_{i-1}$, we use the moment generating function for $X_i$ from (\ref{mgfx})-(\ref{mgfx12}) for the expression of $\mbox{Pr}[I_i + P'_{i} +Y_i< S_{i-1}]$ and $\mbox{Pr}[I_i + P'_{i} +Y_i < W_{i-1} + S_{i-1}]$:
	\begin{align*}
	\mbox{Pr}[I_i + P'_{i} +Y_i< S_{i-1}] &= M_{X}(\mu) \\
	\mbox{Pr}[I_i + P'_{i} +Y_i < W_{i-1} + S_{i-1}] &= M_{X}(\mu)+\mu M_{(X,1)}(\mu),
	\end{align*}
	and $\mbox{Pr}[X_i > W_{i-1} + S_{i-1}] = 1 - \mbox{Pr}[X_i < W_{i-1} + S_{i-1}]$. Then, the stationary probabilities are $p_B=\mbox{Pr}[K_i = (B)]=\frac{M_{X}(\mu)}{1-\mu M_{(X,1)}(\mu)}$ and $p_I = 1- p_B$. 
	
	\subsection{Average AoI}
	
	We next evaluate $\mathbb{E}[XT]$ and put it in (\ref{aaoi}) along with (\ref{ap1}) to get a closed form expression for average AoI. We treat the two conditions $K_{i-1}=(Id)$ and $K_{i-1}=(B)$ separately.
	
	\subsubsection{$\mathbb{E}[X_iT_i \ | \ K_{i-1}=(Id)]$}
	
	In this case, packet $i-1$ finds the second queue in (Id) state. $X_i = I_i + P'_{i-1} +Y_i$ and if $I_i + P'_{i} +Y_i > S_{i-1}$, then $T_i = P'_i + S_i$. If $I_i + P'_{i} +Y_i < S_{i-1}$, then $T_i = P'_i + W_i + S_i$ where $W_i$ is the residual service time observed by packet $i$ before entering the transmission server. We evaluate the conditional expectation as:
	\begin{align*}
	&\mathbb{E}[X_iT_i|K_{i-1}=(Id)] \\&= \mathbb{E}[(I_i + P'_{i-1} +Y_i)(P'_i + S_i)] \\ &\qquad +\mathbb{E}[(I_i + P'_{i-1} +Y_i)W_i \mathbbm{1}_{I_i + P'_{i} +Y_i < S_i}] \\ &= \mathbb{E}[(I_i + P'_{i-1} +Y_i)(P'_i + S_i)] \\&\qquad +\mathbb{E}[(I_i + P'_{i-1} +Y_i)W_ie^{-\mu (I_i + P'_{i} +Y_i)}] \\ &= \mathbb{E}^2[P'] + \frac{\mathbb{E}[P']}{\lambda} + \frac{1}{\lambda \mu} + \frac{\mathbb{E}[P']}{\mu}+\mathbb{E}[Y]\mathbb{E}[P']+\frac{\mathbb{E}[Y]}{\mu}  \\ &\ + \frac{\lambda}{\mu (\lambda + \mu)} M_{P'}(\mu)(\frac{M_{Y}(\mu)}{\lambda+\mu}+\mathbb{E}[P']M_{Y}(\mu)+M_{(Y,1)}(\mu))
	\end{align*}
	where $\mathbb{E}[P']=M_{(P',1)}(\gamma)|_{\gamma=0}=\frac{M_{(P,1)}(\lambda)}{M_{P}(\lambda)}$ and $\mathbb{E}[Y]=M_{(Y,1)}(\gamma)|_{\gamma=0}=\frac{1-M_{P}(\lambda)-\lambda M_{(P,1)}(\lambda)}{\lambda M_{P}(\lambda)}$.

	\subsubsection{$\mathbb{E}[X_iT_i \ | \ K_{i-1}=(B)]$}
	
	In this case, packet $i-1$ finds the second queue in (B) state. $X_i = I_i + P'_{i-1} +Y_i$ and if $I_i + P'_{i} +Y_i > W_{i-1} + S_{i-1}$, then $T_i = P'_i + S_i$. If $I_i + P'_{i} +Y_i < W_{i-1} + S_{i-1}$, then $T_i = P'_i + W_i + S_i$. We evaluate the conditional expectation $\mathbb{E}[X_iT_i|K_{i-1}=(B)]$ in Appendix \ref{App:extMG1P}.
	We finally use ergodicity of the system to get:
	\begin{align*}\nonumber
	\mathbb{E}[X_iT_i]&=\mathbb{E}[X_iT_i|K_{i-1}=(B)]p_{B} + \mathbb{E}[X_iT_i|K_{i-1}=(Id)]p_{I} \end{align*} 
	where $p_B, p_I$ are stationary busy, idle probabilities.
	
	\subsection{Average Peak AoI}
	
	In this subsection, we evaluate $\mathbb{E}[X_{i^*} + T_{i^*}]$ where $i^*$ is the packet index corresponding to the minimum index in a given service period. We have $\mathbb{E}[X_{i^*}+T_{i^*}]=\frac{\mathbb{E}[(X_{i}+T_{i})\mathbbm{1}_{i=i^*}]}{\mbox{Pr}(i=i^*)}$ where $\mathbbm{1}_{i=i^*}$ is the indicator function of whether a given packet is the minimum index in a given service period and $\mbox{Pr}(i=i^*)$ refers to its probability. As before, we will treat the conditions $K_{i-1}=(Id)$ and $K_{i-1}=(B)$ separately for both terms.

	\subsubsection{$\mathbb{E}[(X_{i}+T_{i})\mathbbm{1}_{i=i^*} \ | \ K_{i-1}=(Id)]$}
	
	In this case, if $I_i + P'_{i} +Y_i> S_{i-1}$, then $T_i = P'_i + S_i$. If $I_i + P'_{i} +Y_i < S_{i-1}$, then $T_i = P'_i + W_i + S_i$. Conditioned on (Id) state for packet $i-1$, the next packet index $i$ will be the minimum index among all those being served together with certainty. Therefore, we have $\mbox{Pr}(i=i^* \ | \ K_{i-1}=(Id)) = 1$. Additionally, we have
	\begin{align*}
	&\mathbb{E}[(X_{i}+T_{i})\mathbbm{1}_{i=i^*} \ | \ K_{i-1}=(Id)]  \\ &=  \mathbb{E}[P'_{i-1} + I_i + Y_i+ P'_i + S_i] +\mathbb{E}[W_i \mathbbm{1}_{I_i + P'_{i} +Y_i<S_i}] \\ &=  \mathbb{E}[P'_{i-1} + I_i + Y_i+ P'_i + S_i] +\mathbb{E}[W_ie^{-\mu (I_i + P'_{i} +Y_i)}]  \\ &=2\mathbb{E}[P'] +\mathbb{E}[Y]+ \frac{1}{\lambda} + \frac{1}{\mu} + \frac{\lambda}{\mu(\lambda + \mu)}M_{P'}(\mu)M_{Y}(\mu).
	\end{align*}

	\subsubsection{$\mathbb{E}[(X_{i}+T_{i})\mathbbm{1}_{i=i^*} \ | \ K_{i-1}=(B)]$}
	
	Conditioned on (B) state observed by packet $i-1$, the next packet index $i$ will be the minimum index among all those being served together only if the next packet $i$ arrives after the residual time $W_{i-1}$. Therefore, if $I_i + P'_{i} +Y_i < W_{i-1}$, then $\mathbbm{1}_{i=i^*} = 0$ and we have $\mbox{Pr}(i=i^* \ | \ K_{i-1}=(B)) = 1 - \frac{\lambda}{\lambda + \mu} M_{P'}(\mu)M_{Y}(\mu)$. In this case, if $I_i + P'_{i} +Y_i > W_{i-1} + S_{i-1}$, then $T_i = P'_i + S_i$. If $W_{i-1} < I_i + P'_{i} +Y_i < W_{i-1} + S_{i-1}$, then $T_i = P'_i + W_i + S_i$ and we have
	\begin{align*}
	&\mathbb{E}[(X_{i}+T_{i})\mathbbm{1}_{i=i^*} \ | \ K_{i-1}=(B)] \\&= \mathbb{E}[('P_{i-1} + I_i +Y_i+ P'_i + S_i)(1-e^{-\mu (I_i +Y_i+ P'_i)})] \\ &\qquad +\mathbb{E}[W_i \mu (I_i +Y_i+ P'_i)e^{-\mu (I_i +Y_i+ P'_i)}] \nonumber \\ &=2\mathbb{E}[P'] + \mathbb{E}[Y] - \frac{(\mathbb{E}[P'] + \frac{1}{\mu})\lambda}{\lambda + \mu}M_{P'}(\mu)M_{Y}(\mu) + \frac{1}{\lambda} + \frac{1}{\mu}. 
	\end{align*}
	We use ergodicity of the system to conclude as follows:
	\begin{align*}\nonumber
	\hspace{-0.1in}\mathbb{E}[(X_{i}+T_{i})\mathbbm{1}_{i=i^*}] &= 2\mathbb{E}[P'] + (1- 2p_B)\frac{\lambda M_{P'}(\mu)M_{Y}(\mu)}{\mu(\lambda + \mu)} \\ & \ + \frac{1}{\lambda} + \frac{1}{\mu} - p_B \frac{\mathbb{E}[P'] \lambda}{\lambda + \mu}M_{P'}(\mu)M_{Y}(\mu), \\
	\mbox{Pr}(i=i^*) &= 1 - p_B \frac{\lambda}{\lambda + \mu} M_{P'}(\mu)M_{Y}(\mu)
	\end{align*}
	and finally we have $\mathbb{E}[X_{i^*} + T_{i^*}] = \frac{\mathbb{E}[(X_{i}+T_{i})\mathbbm{1}_{i=i^*}] }{\mbox{Pr}(i=i^*)}$.
	
	\section{Numerical Results}
	\label{sec:Numres}
	
	In this section, we provide numerical results for AoI with respect to system parameters for different service distributions. We also performed packet-based queue simulations offline for $10^6$ packets as verification of all numerical results. We will obtain the best operating point determined by mean service times $\mathbb{E}[P]$ and $\mathbb{E}[S]=\frac{1}{\mu}$ given that $\frac{1}{\mu}=g(\mathbb{E}[P])$ where $g(.)$ is a monotone decreasing function. For simplicity, we use $g(\mathbb{E}[P])=B_0 e^{-\alpha \mathbb{E}[P]}$. While our theoretical development does not depend on the specific $g(.)$ function, this selection of $g(.)$ is smooth and convex which well represents the diminishing returns obtained by enhanced processing of data. We let the expected processing time to be selected from $P_{min} \leq \mathbb{E}[P] \leq P_{max}$. We take $P_{min}=1$ and $P_{max}=10$.
	
	We use Gamma distributed computation time with mean $\mathbb{E}[P]$. In particular, we use the probability density function $f_P(p)=\frac{k^k \kappa^k}{\Gamma(k)}p^{k-1}e^{-k \kappa p}$ for $p \geq 0$ where $\kappa=\frac{1}{\mathbb{E}[P]}$ and $k >0$ determines the variance. The variance gets larger as $k$ gets smaller. We have the following closed form expressions for this Gamma distribution: 
	\begin{align*}
  	M_{P}(\gamma)&=\left(1+\frac{\gamma }{k \kappa}\right)^{-k}, \ M_{(P,1)}(\gamma)=\frac{1}{\kappa}\left(1+\frac{\gamma }{k \kappa}\right)^{-k-1}, \\ &MGF_{(P,1)}(\gamma)=\frac{k+1}{k \kappa}\left(1+\frac{\gamma }{k \kappa}\right)^{-k-2}.
	\end{align*}

\begin{figure}[!t]
		\centering{
			\hspace{-0.3cm} 
			\includegraphics[totalheight=0.35\textwidth]{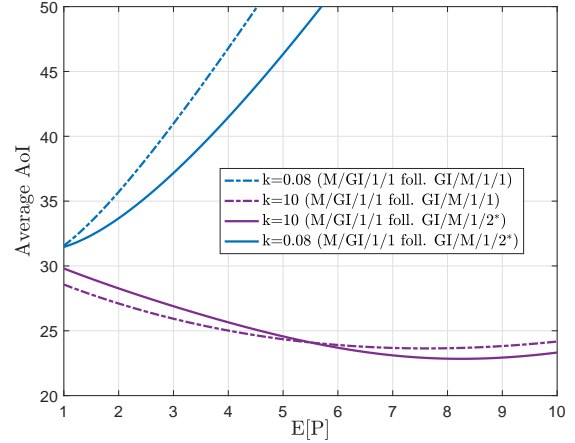}}\vspace{-0.15in}
		\caption{\sl Average AoI with respect to $\mathbb{E}[P]$ for fixed $\lambda=0.4$, $B_0=15$ and $\alpha=0.1$ comparing M/GI/1/1 followed by GI/M/1/1 scheme and M/GI/1/1 followed by GI/M/1/$2^*$ scheme.}\vspace{-0.25in}
		\label{fig:comp2} 
	\end{figure}

\begin{figure}[!t]
		\centering{
			\hspace{-0.3cm} 
			\includegraphics[totalheight=0.35\textwidth]{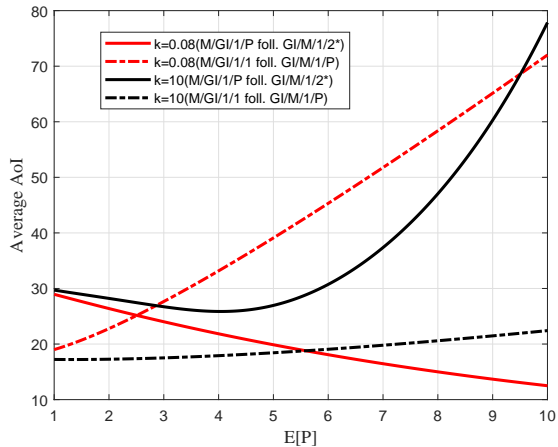}}\vspace{-0.15in}
		\caption{\sl Average AoI with respect to $\mathbb{E}[P]$ for fixed $\lambda=0.4$, $B_0=15$ and $\alpha=0.1$ comparing M/GI/1 preemption followed by GI/M/1/1 and M/GI/1/1 followed by GI/M/1 preemption.}\vspace{-0.2in}
		\label{fig:comp1} 
	\end{figure}

	We start with Figs. \ref{fig:comp2} and \ref{fig:comp1} where we plot average AoI with respect to $\mathbb{E}[P]$ for all four schemes under different computing time variances. We observe in Fig. \ref{fig:comp1} that for non-preemptive schemes average AoI decreases uniformly as the variance of computing time is decreased. We observe this trend for average peak AoI as well in our extended numerical studies that are left outside of the current paper. These observations support the usefulness of determinacy in this tandem queue system with non-preemptive policies. On the contrary, we observe just the opposite behavior in Fig. \ref{fig:comp1} where larger variance of computing distribution yields smaller average AoI for preemptive policies. Our observations contrasting the preemptive and non-preemptive schemes are in line with those in the seminal paper \cite{talak2018can} for single-server first come first served systems. Our work extends \cite{talak2018can} at least numerically in the tandem queue model with packet management and warrant further research in this curious phenomenon on the variance of arrival/service distributions and relations to AoI. It is also remarkable that preemptive schemes have smaller average AoI than non-preemptive ones. For smaller $\mathbb{E}[P]$ and smaller variance, preemption in the second queue yields smaller average AoI whereas for larger $\mathbb{E}[P]$ and larger variance, preemption in the first queue is dominant. 
	
	\begin{figure}[!t]
		\centering{
			\hspace{-0.3cm} 
			\includegraphics[totalheight=0.35\textwidth]{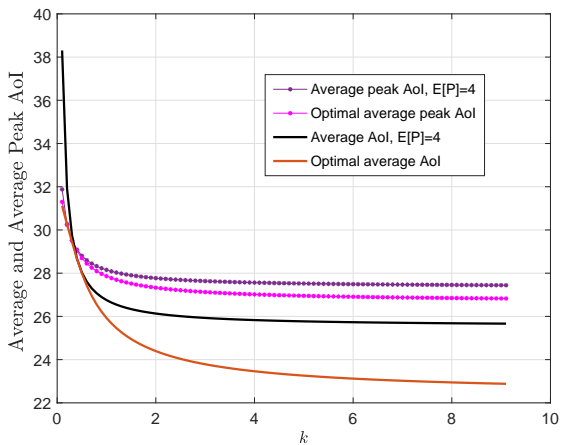}}\vspace{-0.15in}
		\caption{\sl Average and average peak AoI with respect to $k$ for fixed $\lambda=0.4$, $B_0=15$ and $\alpha=0.1$ in M/GI/1/1 followed by GI/M/1/$2^*$.}\vspace{-0.05in}
		\label{fig:numGM12vsk} 
	\end{figure}
	
\begin{figure}[!t]
		\centering{
			\hspace{-0.3cm} 
			\includegraphics[totalheight=0.35\textwidth]{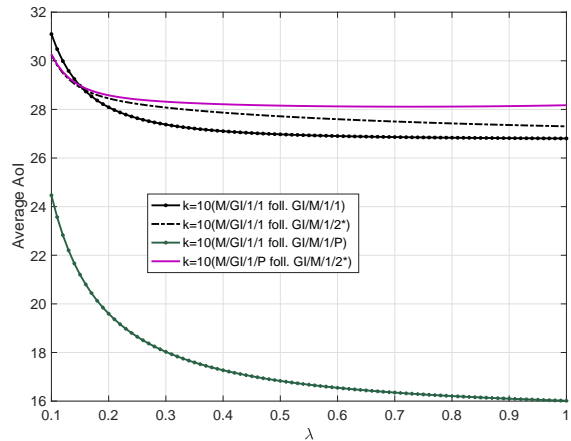}}\vspace{-0.15in}
		\caption{\sl Average AoI with respect to $\lambda$ for fixed $\mathbb{E}[P]=2$, $k=10$, $B_0=15$ and $\alpha=0.1$ comparing all four schemes.}\vspace{-0.05in}
		\label{fig:compp} 
	\end{figure}

	Next, in Fig. \ref{fig:numGM12vsk}, we compare the gains in average AoI by judiciously selecting the operating point $\mathbb{E}[P]$. In particular, we select $\mathbb{E}[P]=4$ arbitrarily and compare its performance with optimal selection. The improvement in average AoI is significantly higher for larger variances while it is not the case for peak AoI. This is analogous to the effect of waiting as in \cite{infocom_w, infocom_arxiv} where larger variance in the service time distribution yields a higher improvement in average AoI. In particular, the mean service time $\mathbb{E}[P]$ has an analogous role as \textit{waiting time} from the point of view of the second queue. Additionally, the improvement in average AoI shows a larger margin compared to average peak AoI. In Fig. \ref{fig:compp}, we provide an instance of average AoI performances of all schemes plotted with respect to arrival rate $\lambda$. We observe that the order between non-preemptive and preemptive schemes change with $\lambda$ while we have seen consistently in our numerical study that preemption in the transmission queue has a significantly better average AoI performance for moderate to large $\lambda$. This comes with increased peak AoI as we will observe in the following numerical results.	
			
	\begin{figure}[!t]
		\centering{
			\hspace{-0.3cm} 
			\includegraphics[totalheight=0.34\textwidth]{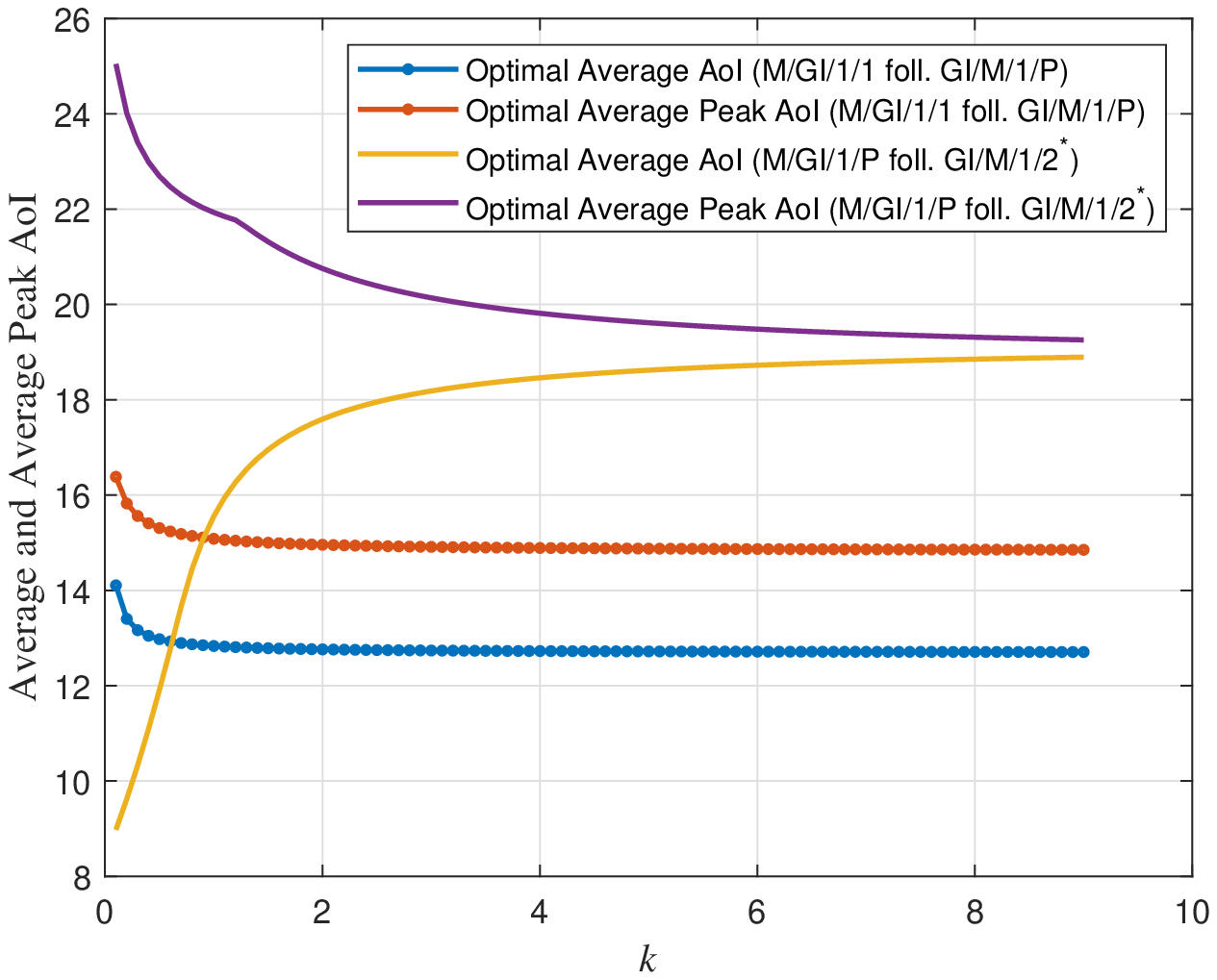}}\vspace{-0.15in}
		\caption{\sl Optimal average AoI and average peak AoI with respect to $k$ for fixed $\lambda=0.4$, $B_0=10$ and $\alpha=0.1$ comparing M/GI/1 with preemption followed by GI/M/1/$2^*$ and M/GI/1/1 followed by GI/M/1 with preemption schemes.}\vspace{-0.15in}
		\label{fig:numMG1Pvsk} 
	\end{figure}
	
	\begin{figure}[!t]
		\centering{
			\hspace{-0.3cm} 
			\includegraphics[totalheight=0.34\textwidth]{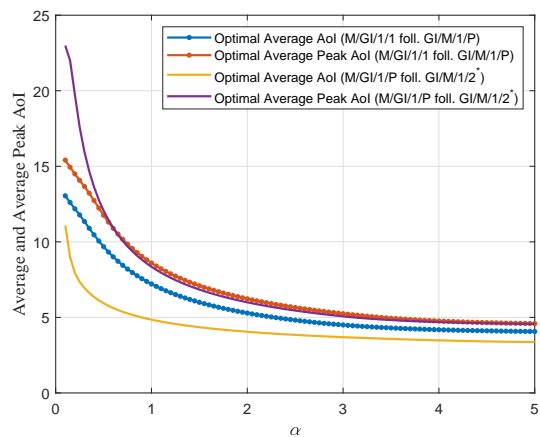}}\vspace{-0.1in}
		\caption{\sl Optimal average AoI and average peak AoI with respect to $\alpha$ for fixed $\lambda=0.4$, $B_0=10$ and $k=0.4$ comparing M/GI/1 with preemption followed by GI/M/1/$2^*$ and M/GI/1/1 followed by GI/M/1 with preemption schemes.}\vspace{-0.1in}
		\label{fig:numGM11vsa} 
	\end{figure}

Next, we plot optimal average AoI and average peak AoI with respect to $k$ (indicator of the variance of $P$) in Fig. \ref{fig:numMG1Pvsk} and with respect to $\alpha$ (the exponent appearing in $g(.)$ function) in Fig. \ref{fig:numGM11vsa} for preemptive schemes. First, note that the larger $\alpha$ is, the larger the reduction in service time of the transmission queue is. We observe monotonicity with respect to $\alpha$ and different points of convergence for different $k$ values. Figs. \ref{fig:numGM11vsa} and \ref{fig:numMG1Pvsk} together show that preemption in the first queue has superior average AoI performance for small $k$ (large computation time variance) at the cost of an average peak AoI penalty whereas preemption in the second queue outperforms the other both in average AoI and average peak AoI for large $k$ (small computation time variance). 

\begin{figure}[!t]
\centering{
\hspace{-0.3cm} 
\includegraphics[totalheight=0.27\textheight]{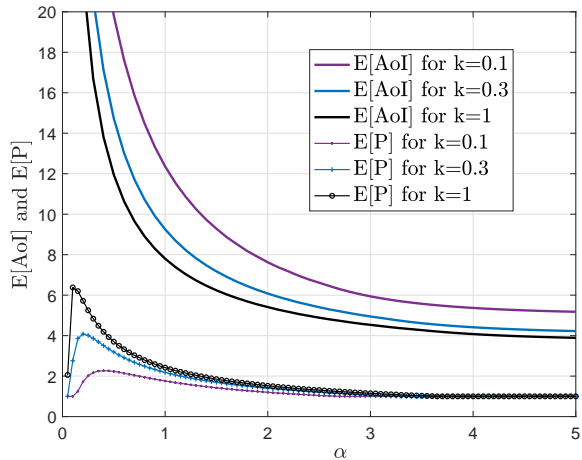}}\vspace{-0.1in}
\caption{\sl Optimal average AoI and optimal $\mathbb{E}[P]$ with respect to $\alpha$ for various $k$ and fixed $\lambda=0.4$ and $B_0=15$ in M/GI/1/1 followed by GI/M/1/$2^*$.}\vspace{-0.1in}
\label{fig:num5} 
\end{figure}

In Fig. \ref{fig:num5}, we investigate the characteristic of best selection of computation time $\mathbb{E}[P]$ with respect to the exponent $\alpha$. We observe that optimizer $E[P]$ has a unimodal shape with respect to $\alpha$ for M/GI/1/1 followed by GI/M/1/$2^*$ and our other offline numerical study shows this unimodal shape in other non-preemptive and preemptive schemes. This unimodal shape suggests that computation has to be ramped up under two extremes when increased computation time cannot provide sufficient time improvement in the subsequent transmission server (small $\alpha$) or when a small computation effort is sufficient to improve the service time in the subsequent transmission server (large $\alpha$). It is also remarkable that there is an inverse relation between the variance of computation time and optimizer $\mathbb{E}[P]$ as well as the optimal average AoI.
		
\begin{figure}[!t]
		\centering{
			\hspace{-0.3cm} 
			\includegraphics[totalheight=0.34\textwidth]{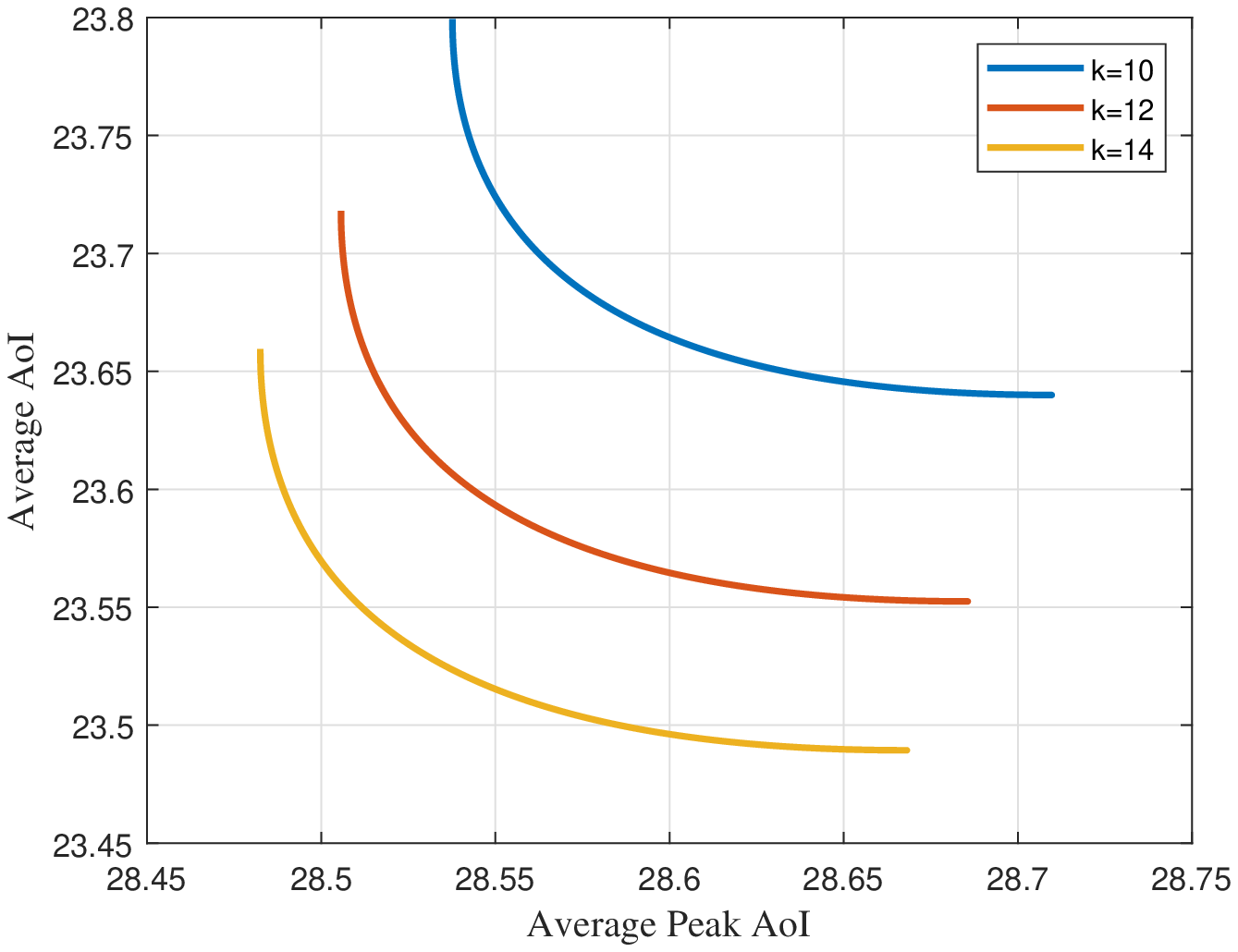}}\vspace{-0.05in}
		\caption{\sl Optimal tradeoff curves for average AoI vs. average peak AoI with differing variances and fixed $\lambda=0.4$, $B_0=15$ and $\alpha=0.1$ in M/GI/1/1 followed by GI/M/1/1.}\vspace{-0.05in}
		\label{fig:numtradeGM11} 
	\end{figure}

\begin{figure}[!t]
	\centering{
		\hspace{-0.3cm} 
		\includegraphics[totalheight=0.34\textwidth]{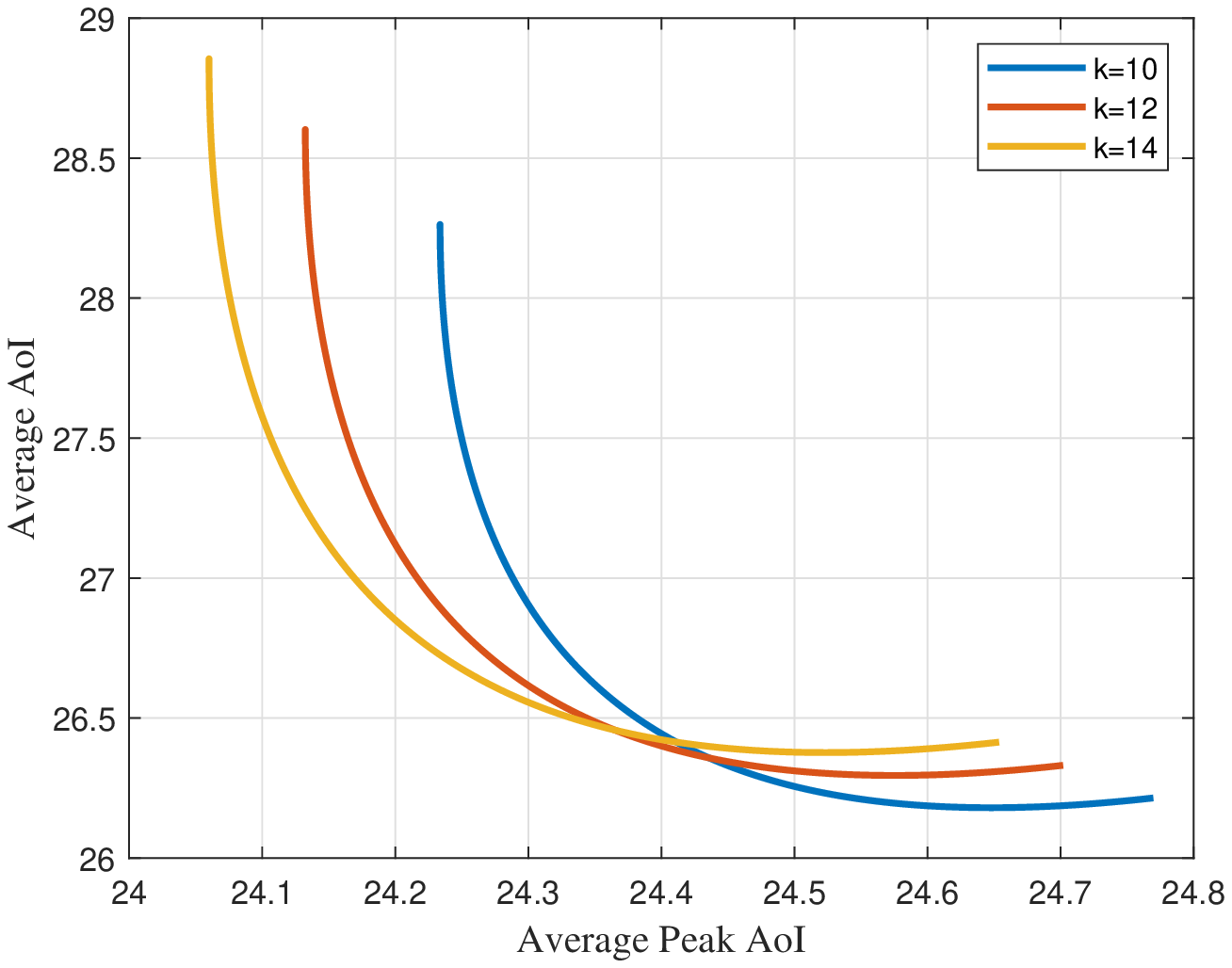}}\vspace{-0.05in}
	\caption{\sl Optimal tradeoff curves for average AoI vs. average peak AoI with differing variances and fixed $\lambda=0.4$, $B_0=15$ and $\alpha=0.1$ in M/GI/1 with preemption followed by GI/M/1/$2^*$.}\vspace{-0.05in}
	\label{fig:numtradeMG1P} 
\end{figure}
	
	\begin{figure}[!t]
		\centering{
			\hspace{-0.3cm} 
			\includegraphics[totalheight=0.34\textwidth]{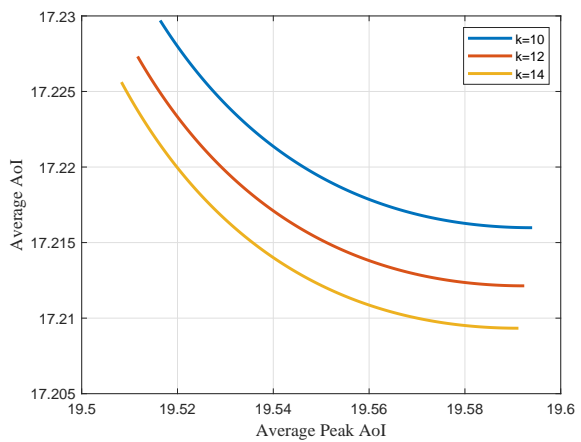}}\vspace{-0.05in}
		\caption{\sl Optimal tradeoff curves for average AoI vs. average peak AoI with differing variances and fixed $\lambda=0.4$, $B_0=15$ and $\alpha=0.1$ in M/GI/1/1 followed by GI/M/1 with preemption.}\vspace{-0.05in}
		\label{fig:numtradeGM1P} 
	\end{figure}

We observe that the optimal values of $\mathbb{E}[P]$ could be quite different for average AoI and average peak AoI. In general a decrease in average AoI comes at the cost of increased average peak AoI. To understand the tradeoff between average AoI and average peak AoI, we optimize weighted sum of AoI and average peak AoI for different weights over the mean computation time.
	\begin{align}\label{ths}
	\min_{\mathbb{E}[P] \geq 0} \omega_1 \mathbb{E}[\Delta] + \omega_2 \mathbb{E}[PAoI]
	\end{align}
	 The expected processing time is in between $P_{min}=1$ and $P_{max}=10$. We dedicate the final three plots in Figs. \ref{fig:numtradeGM11} - \ref{fig:numtradeGM1P} for the tradeoff between average AoI and average peak AoI. In these figures, we plot the optimal tradeoff obtained by solving the weighted optimization in (\ref{ths}) for differing service time variances. In particular, for each $k$ determining the service time variance, we solve (\ref{ths}) for all possible $\omega_1$ and $\omega_2$ and plot all possible operating points as tuples of average AoI and average peak AoI. We observe that this tradeoff becomes more apparent for smaller service time variances in general. For non-preemptive M/GI/1/1 followed by GI/M/1/1, the plot is as in Fig. \ref{fig:numtradeGM11} whereas for preemptive schemes we observe Figs. \ref{fig:numtradeMG1P}, \ref{fig:numtradeGM1P}. We observe that preemption in the second queue outperforms the rest in both average AoI and average peak AoI. The other two schemes beat one another in average AoI but has the opposite order for average peak AoI. Finally, we note that these tradeoff curves take trivial forms as variance is increased ($k$ is decreased). In particular, the plots become a single point in the plane for M/GI/1/1 followed by GI/M/1/1 and M/GI/1/1 followed by GI/M/1 with preemption. It becomes two extreme points in the plane for M/GI/1 with preemption followed by GI/M/1/$2^*$. We do not add these extreme points in these plots as they reside far from the shown plots. \vspace{-0.2in}

	\section{Conclusions}
	\label{sec:Conc}
	
	In this paper, we consider stationary distribution analysis for average AoI and average peak AoI in a tandem non-preemptive queue with a computation server and a transmission server whose arrivals come from the computation queue. Our analysis covers various packet management schemes at the transmission queue with potential preemption allowed at the computation or the transmission. Additionally, we assume a functional dependence between the mean service times of the first and the second queues. We obtain closed form expressions for average AoI and average peak AoI in this system. Our expressions provide explicit relations among the parameters in the system. Our numerical results show the advantages obtained by judiciously determining the operating point in the tradeoff achieved between average AoI and average peak AoI for different computation time distributions. \vspace{-0.2in}

	\appendix
	
	\vspace{-0.1in}
	
	\subsection{$M_{(P',1)}(\gamma)$, $M_{(Y,1)}(\gamma)$ and $M_{(Y,2)}(\gamma)$ for  M/GI/1 with Preemption followed by GI/M/1/$2^*$}\label{App:extMGFp1MGFy1}	
	\begin{align*}
M_{(P',1)}(\gamma)&=-\frac{\mathrm{d}M_{P'}(\gamma)}{\mathrm{d}\gamma}=-\frac{1}{M_{P}(\lambda)}\frac{\mathrm{d}M_{P}(\gamma+\lambda)}{\mathrm{d}\gamma}\\&=\frac{M_{(P,1)}(\gamma+\lambda)}{M_{P}(\lambda)}.\end{align*}
\begin{align*}
&M_{(Y,1)}(\gamma)=-\frac{\mathrm{d}}{\mathrm{d}\gamma}M_{Y}(\gamma)=-\frac{\mathrm{d}}{\mathrm{d}\gamma}\frac{(\lambda+\gamma)M_{P}(\lambda)}{\gamma+\lambda M_P(\gamma+\lambda)}\\
	&=\frac{(\lambda+\gamma)M_P(\lambda)(1-\lambda M_{(P,1)}(\lambda+\gamma))}{(\gamma+\lambda M_P(\lambda+\gamma))^2} \\ 
	&\hspace{1in} -\frac{M_P(\lambda)(\gamma+\lambda M_P(\lambda+\gamma))}{(\gamma+\lambda M_P(\lambda+\gamma))^2}\\
	&=\frac{\lambda M_P(\lambda)(1-(\lambda+\gamma)M_{(P,1)}(\lambda+\gamma)-M_P(\lambda+\gamma))}{(\gamma+\lambda M_P(\lambda+\gamma))^2}.
\end{align*}
\begin{align*}
&M_{(Y,2)}(\gamma)=-\frac{\mathrm{d}M_{(Y,1)}(\gamma)}{\mathrm{d}\gamma} \\
	&=\frac{(-\lambda^{2}-\lambda\gamma)(\gamma+\lambda M_P(\lambda+\gamma))M_P(\lambda)M_{(P,2)}(\lambda+\gamma)}{(\gamma+\lambda M_P(\lambda+\gamma))^3}\\
	&\ +\frac{2\lambda M_P(\lambda)(1-(\lambda+\gamma)M_{(P,1)}(\lambda+\gamma)-M_P(\lambda+\gamma))}{(\gamma+\lambda M_P(\lambda+\gamma))^3} \\
	&\ -\frac{2\lambda M_{(P,1)}(\lambda+\gamma)\lambda M_P(\lambda)(1-M_P(\lambda+\gamma))}{(\gamma+\lambda M_P(\lambda+\gamma))^3} \\
	&\ +\frac{2\lambda M_{(P,1)}(\lambda+\gamma)\lambda M_P(\lambda)(\lambda+\gamma)M_{(P,1)}(\lambda+\gamma)}{(\gamma+\lambda M_P(\lambda+\gamma))^3}.
	\end{align*}	
	\subsection{$\mathbb{E}[X_iT_i|K_{i-1}=(B)]$ for  M/GI/1 with Preemption followed by GI/M/1/$2^*$}\label{App:extMG1P}	
	\begin{align*}\nonumber
	\hspace{-0.0in}&\mathbb{E}[X_iT_i|K_{i-1}=(B)]  = \mathbb{E}[(I_i + P'_{i-1} +Y_i)(P'_i + S_i)] \\&\ +\mathbb{E}[(I_i + P'_{i-1} +Y_i)W_ie^{-\mu (I_i + P'_{i} +Y_i)}] \\  &\ + \mathbb{E}[(I_i + P'_{i-1} +Y_i)W_i\mu (I_i + P'_{i} +Y_i) e^{-\mu (I_i + P'_{i} +Y_i)}] \\ \nonumber &= \mathbb{E}^2[P'] + \frac{\mathbb{E}[P']}{\lambda} + \frac{1}{\lambda \mu} + \frac{\mathbb{E}[P']}{\mu}+\mathbb{E}[Y]\mathbb{E}[P']+\frac{\mathbb{E}[Y]}{\mu}  \\ \nonumber &\ + \frac{\lambda}{\mu (\lambda + \mu)} M_{P'}(\mu)(\frac{M_{Y}(\mu)}{\lambda+\mu}+\mathbb{E}[P']M_{Y}(\mu)+M_{Y,1}(\mu)) \end{align*} \begin{align*} &\ +\frac{2\lambda}{(\lambda+\mu)^3}M_{P'}(\mu)M_{Y}(\mu)+\frac{\lambda}{(\lambda+\mu)^2}M_{(P',1)}(\mu)M_{Y}(\mu)\\&\ +\frac{\lambda}{(\lambda+\mu)^2}M_{P'}(\mu)M_{(Y,1)}(\mu) +\frac{\lambda\mathbb{E}[P']}{(\lambda+\mu)^2}M_{P'}(\mu)M_{Y}(\mu)\\&\ +\frac{\lambda\mathbb{E}[P']}{\lambda+\mu}M_{(P',1)}(\mu)M_{Y}(\mu)+\frac{\lambda\mathbb{E}[P']}{\lambda+\mu}M_{P'}(\mu)M_{(Y,1)}(\mu)\\ &\ +\frac{\lambda}{(\lambda+\mu)^2}M_{P'}(\mu)M_{(Y,1)}(\mu)+\frac{\lambda}{\lambda+\mu}M_{(P',1)}(\mu)M_{(Y,1)}(\mu)\\&\ +\frac{\lambda}{\lambda+\mu}M_{P'}(\mu)M_{(Y,2)}(\mu). 
	\end{align*}

\subsection{$\mathbb{E}[W']$ and $\mathbb{E}[S']$ for  M/GI/1/1 followed by GI/M/1 with Preemption}\label{App:extGM1P}

	Let us denote $X'_{j}$ as independent realizations of $X$ conditioned on the event $\{X<S\}$ where $S$ is a rate $\mu$ exponential random variable independent of $X$. We use $W'$ to denote the waiting time of a packet before taken into service in the equivalent queuing model. Since $W'=\sum_{j=0}^{B}X'_{j}$ and $B$ is geometric with success probability $\mathbb{P}[X>S]$, we have:
	\begin{align*}
	\mathbb{E}[W']&=\frac{1-\mathbb{P}[X>S]}{\mathbb{P}[X>S]}\mathbb{E}[X'],\\
	\mathbb{P}[X>S]&=1-\mathbb{P}[X<S]=1-\frac{\lambda}{\lambda+\mu}M_{P}(\mu).
	\end{align*}
	From (\ref{lemma1}), we have:
	\begin{align*}
	M_{X'}(\gamma)&=\frac{M_{X}(\gamma+\mu)}{M_{X}(\mu)}\\&=\frac{\frac{\lambda}{\lambda+\mu+\gamma}M_P(\mu+\gamma)}{\frac{\lambda}{\lambda+\mu}M_{P}(\mu)}=\frac{(\lambda+\mu)M_P(\mu+\gamma)}{(\lambda+\mu+\gamma)M_P(\mu)},\\
	\mathbb{E}[X']&=M_{(X',1)}(\gamma)|_{\gamma=0}=\frac{(\lambda+\mu)M_{(P,1)}(\mu)+M_{P}(\mu)}{(\lambda+\mu)M_{P}(\mu)}.
	\end{align*}
	Now we have $\mathbb{E}[W']$ as:
	\begin{align*}
	\mathbb{E}[W']&=\frac{1-\mathbb{P}[X>S]}{\mathbb{P}[X>S]}\mathbb{E}[X']\\&=\frac{\frac{\lambda}{\lambda+\mu}M_{P}(\mu)}{1-\frac{\lambda}{\lambda+\mu}M_P(\mu)}\frac{(\lambda+\mu)M_{(P,1)}(\mu)+M_{P}(\mu)}{(\lambda+\mu)M_{P}(\mu)}\\
	&=\frac{\lambda(\lambda+\mu)M_{(P,1)}(\mu)+\lambda M_{P}(\mu)}{(\lambda+\mu)(\lambda+\mu-\lambda M_{P}(\mu))}.
	\end{align*}
	Then we consider $\mathbb{E}[S']$. $S_i'$ is a rate $\mu$ exponential random variable conditioned on the event $\{S<X\}$ where $X$ is general distribution random variable. From (\ref{lemma2}), we have:
	\begin{align*}
	M_{S'}(\gamma)&=\frac{1}{1-M_{X}(\mu)}(\frac{\mu}{\mu+\gamma}-\frac{\mu M_X(\gamma+\mu)}{\mu+\gamma})\\&=\frac{1}{1-\frac{\lambda}{\lambda+\mu}M_{P}(\mu)}(\frac{\mu}{\mu+\gamma}-\frac{\mu \frac{\lambda}{\lambda+\mu+\gamma}M_P(\mu+\gamma)}{\mu+\gamma})\\
	&=\frac{\lambda+\mu}{\lambda+\mu-\lambda M_{P}(\mu)}\frac{\mu}{\mu+\gamma}(1-\frac{\lambda M_P(\mu+\gamma)}{\lambda+\mu+\gamma}),\\
	\mathbb{E}[S']&=M_{(S',1)}(\gamma)|_{\gamma=0}\\&=\frac{1}{\mu}-\frac{\lambda}{\lambda+\mu-\lambda M_{P}(\mu)}(M_{(P,1)}(\mu)+\frac{M_P(\mu)}{\lambda+\mu}).
	\end{align*}


\begin{thebibliography}{10}
		
		\bibitem{pimrc19}
		P.~Zou, O.~Ozel, and S.~Subramaniam.
		\newblock Trading off computation with transmission in status update systems.
		\newblock In {\em IEEE PIMRC}, September 2019.
		
		\bibitem{aydin2018}
		A.~Chopra, H.~Aydin, S.~Rafatirad, and H.~Homayoun.
		\newblock Optimal allocation of computation and communication in an {IoT}
		network.
		\newblock {\em ACM Transactions on Design Automation of Electronic Systems},
		23(6):78, 2018.
		
		\bibitem{opadere2019joint}
		J.~Opadere, Q.~Liu, N.~Zhang, and T.~Han.
		\newblock Joint computation and communication resource allocation for
		energy-efficient mobile edge networks.
		\newblock In {\em IEEE ICC}, 2019.
		
		\bibitem{kaul2012real}
		S.~Kaul, R.~Yates, and M.~Gruteser.
		\newblock Real-time status: How often should one update ?
		\newblock In {\em INFOCOM}, pages 2731--2735. IEEE, 2012.
		
		\bibitem{kaul2012status}
		S.K. Kaul, R.D. Yates, and M.~Gruteser.
		\newblock Status updates through queues.
		\newblock In {\em Information Sciences and Systems (CISS), 2012 46th Annual
			Conference on}, pages 1--6. IEEE, 2012.
		
		\bibitem{inoue2018general}
		Y.~Inoue, H.~Masuyama, T.~Takine, and T.~Tanaka.
		\newblock A general formula for the stationary distribution of the age of
		information and its application to single-server queues.
		\newblock {\em arXiv preprint arXiv:1804.06139}, 2018.
		
		\bibitem{costa2016age}
		M.~Costa, M.~Codreanu, and A.~Ephremides.
		\newblock On the age of information in status update systems with packet
		management.
		\newblock {\em IEEE Transactions on Information Theory}, 62(4):1897--1910,
		2016.
		
		\bibitem{kam2018age}
		C.~Kam, S.~Kompella, G.D. Nguyen, J.E. Wieselthier, and A.~Ephremides.
		\newblock On the age of information with packet deadlines.
		\newblock {\em IEEE Transactions on Information Theory}, 2018.
		
		\bibitem{najm2016age}
		E.~Najm and R.~Nasser.
		\newblock Age of information: The gamma awakening.
		\newblock In {\em Information Theory (ISIT), 2016 IEEE International Symposium
			on}, pages 2574--2578. Ieee, 2016.
		
		\bibitem{2018information}
		A.~Baknina, O.~Ozel, J.~Yang, S.~Ulukus, and A.~Yener.
		\newblock Sending information through status updates.
		\newblock In {\em IEEE ISIT}, 2018.
		
		\bibitem{bacinoglu2015age}
		B.~T. Bacinoglu, E.~T. Ceran, and E.~Uysal-Biyikoglu.
		\newblock Age of information under energy replenishment constraints.
		\newblock In {\em USCD ITA}, February 2015.
		
		\bibitem{yates2015lazy}
		R.~Yates.
		\newblock Lazy is timely: Status updates by an energy harvesting source.
		\newblock In {\em IEEE ISIT}, June 2015.
		
		\bibitem{wu2017optimal_ieee}
		X.~Wu, J.~Yang, and J.~Wu.
		\newblock Optimal status update for age of information minimization with an
		energy harvesting source.
		\newblock {\em IEEE Trans. on Green Communications and Networking}, 2(1), March
		2018.
		
		\bibitem{arafa2017age}
		A.~Arafa and S.~Ulukus.
		\newblock Age-minimal transmission in energy harvesting two-hop networks.
		\newblock In {\em IEEE Globecom}, December 2017.
		
		\bibitem{bacinoglu2017scheduling}
		B.T. Bacinoglu and E.~Uysal-Biyikoglu.
		\newblock Scheduling status updates to minimize age of information with an
		energy harvesting sensor.
		\newblock In {\em IEEE ISIT}, pages 1122--1126. IEEE, 2017.
		
		\bibitem{farazi2018average}
		S.~Farazi, A.G. Klein, and D.R. Brown.
		\newblock Average age of information for status update systems with an energy
		harvesting server.
		\newblock In {\em IEEE INFOCOM WORKSHPS}, pages 112--117, 2018.
		
		\bibitem{bedewy2017age}
		A.~M. Bedewy, Y.~Sun, and N.~B. Shroff.
		\newblock Age-optimal information updates in multihop networks.
		\newblock {\em Available at arXiv:1701.05711}, 2017.
		
		\bibitem{talak2017minimizing}
		R.~Talak, S.~Karaman, and E.~Modiano.
		\newblock Minimizing age-of-information in multi-hop wireless networks.
		\newblock In {\em Communication, Control, and Computing (Allerton), 2017 55th
			Annual Allerton Conference on}, pages 486--493. IEEE, 2017.
		
		\bibitem{yates2018age}
		R.D. Yates.
		\newblock The age of information in networks: Moments, distributions, and
		sampling.
		\newblock {\em arXiv preprint arXiv:1806.03487}, 2018.
		
		\bibitem{yates2018status}
		R.D. Yates.
		\newblock Status updates through networks of parallel servers.
		\newblock In {\em 2018 IEEE International Symposium on Information Theory
			(ISIT)}, pages 2281--2285. IEEE, 2018.
		
		\bibitem{maatouk2018age}
		A.~Maatouk, M.~Assaad, and A.~Ephremides.
		\newblock The age of updates in a simple relay network.
		\newblock {\em arXiv preprint arXiv:1805.11720}, 2018.
		
		\bibitem{Alabbasi2018JointIF}
		A.~Alabbasi and V.~Aggarwal.
		\newblock Joint information freshness and completion time optimization for
		vehicular networks.
		\newblock {\em CoRR}, abs/1811.12924, 2018.
		
		\bibitem{Gong2019ReducingAF}
		J.~Gong, Q.~Kuang, X.~Chen, and X.~Ma.
		\newblock Reducing age-of-information for computation-intensive messages via
		packet replacement.
		\newblock {\em CoRR}, abs/1901.04654, 2019.
		
		\bibitem{xu2019peak}
		C.~Xu, H.~H. Yang, X.~Wang, and T.Q.S Quek.
		\newblock Optimizing information freshness in computing enabled {IoT} networks.
		\newblock {\em arXiv preprint arXiv:1910.05578}, 2019.
		
		\bibitem{infocom_w}
		P.~Zou, O.~Ozel, and S.~Subramaniam.
		\newblock On the benefits of waiting in status update systems.
		\newblock In {\em IEEE INFOCOM WORKSHPS}, April 2019.
		
		\bibitem{infocom_arxiv}
		P.~Zou, O.~Ozel, and S.~Subramaniam.
		\newblock Waiting before serving: A companion to packet management in status
		update systems.
		\newblock {\em arXiv preprint arXiv:1901.02873}, 2019.
		
		\bibitem{talak2018can}
		R.~Talak, S.~Karaman, and E.~Modiano.
		\newblock Can determinacy minimize age of information?
		\newblock {\em arXiv preprint arXiv:1810.04371}, 2018.
		
		\bibitem{isit_arxiv}
		P.~Zou, O.~Ozel, and S.~Subramaniam.
		\newblock Relative age of information: A new metric for status update systems.
		\newblock In {\em IEEE ITW}, August 2019.
		
	\end{thebibliography}
\end{document}